\pdfoutput=1
\PassOptionsToPackage{hyphens}{url}
\documentclass[letterpaper,twocolumn,10pt]{article}
\usepackage{usenix-2020-09}
\usepackage{tikz}
\usetikzlibrary{positioning,arrows.meta,calc,patterns,decorations.pathreplacing,fit,backgrounds}
\newcommand{\figfontS}{\fontfamily{Crlt-LF}\fontsize{8}{9.2}\selectfont}
\definecolor{kbRed}{HTML}{D85A4D}
\definecolor{kbRedL}{HTML}{F2BFB7}
\definecolor{kbBlue}{HTML}{3D6FA0}
\definecolor{kbGreen}{HTML}{4A9D7A}
\definecolor{kbGold}{HTML}{D9A65A}
\definecolor{kbViolet}{HTML}{8E72B5}
\definecolor{kbGray}{HTML}{7F7F7F}

\definecolor{ppBlue}{HTML}{B4C7E7}
\definecolor{ppViolet}{HTML}{D3C6EA}
\definecolor{ppGreen}{HTML}{C5E0B4}
\definecolor{ppTeal}{HTML}{AFDDD5}
\definecolor{ppDarkRed}{HTML}{C00000}
\definecolor{apInk}{HTML}{000000}
\definecolor{apEdge}{HTML}{595959}
\definecolor{apStone}{HTML}{7F7F7F}
\definecolor{apStoneL}{HTML}{EDEDED}
\definecolor{apCache}{HTML}{E2F3EE}
\definecolor{apCacheD}{HTML}{BCE3D8}
\definecolor{apLimb}{HTML}{7E57C2}
\definecolor{apLimbL}{HTML}{E4DCF2}
\definecolor{apPetrol}{HTML}{2F5597}
\definecolor{apPetrolL}{HTML}{DAE3F3}
\definecolor{apBrick}{HTML}{C00000}
\definecolor{apBrickL}{HTML}{F4CCCC}
\definecolor{apCore}{HTML}{DEEBF7}
\tikzset{
  dband/.style={draw=none, fill=#1},
  dbandlab/.style={font=\figfontS, text=black, align=center, inner sep=0.5pt},
  dbox/.style={draw=apEdge, line width=0.5pt, rounded corners=0.8pt, font=\figfontS, text=black, align=center,
               inner sep=1.4pt, fill=white},
  dlab/.style={font=\figfontS, text=black, inner sep=0.8pt, align=center},
  dgo/.style={-{Latex[length=1.6mm,width=1.3mm]}, line width=0.8pt, draw=apPetrol},
  dmem/.style={-{Latex[length=1.6mm,width=1.3mm]}, line width=0.7pt, draw=apBrick, dash pattern=on 2pt off 1.2pt},
  dplain/.style={-{Latex[length=1.5mm,width=1.2mm]}, line width=0.6pt, draw=apEdge},
  dtile/.style={draw=apEdge, line width=0.45pt},
  dgrid/.style={draw=apStone!70, line width=0.2pt},
}

\usepackage{xurl}
\hypersetup{breaklinks=true}
\usepackage{epsfig,graphicx,endnotes,subfig,multirow,amsmath,amssymb,algorithm,algpseudocode}
\usepackage{xcolor,xspace,booktabs}
\usepackage{balance}
\usepackage{pifont}
\usepackage{colortbl}
\usepackage{enumitem}
\usepackage{flushend}
\usepackage[T1]{fontenc}
\usepackage{color,soul}
\usepackage{threeparttable}
\usepackage{placeins}
\newcommand{\cmark}{\textcolor{green!60!black}{\ding{51}}}
\newcommand{\xmark}{\textcolor{red!80!black}{\ding{55}}}

\tikzset{
  apnode/.style={draw=black!55, rounded corners=1.2pt, font=\figfontS, align=center,
                 inner sep=1.6pt, minimum height=4.6mm, fill=white},
  apsys/.style={apnode, fill=kbRedL},
  apbase/.style={apnode, fill=kbBlue!18},
  apdata/.style={apnode, fill=black!6},
  aparrow/.style={-{Stealth[length=1.5mm,width=1.3mm]}, line width=0.7pt, draw=black!70},
  apred/.style={-{Stealth[length=1.5mm,width=1.3mm]}, line width=0.75pt, draw=kbRed},
  apblue/.style={-{Stealth[length=1.5mm,width=1.3mm]}, line width=0.75pt, draw=kbBlue},
  aplabel/.style={font=\figfontS, inner sep=0.8pt, align=center},
  apstep/.style={circle, draw=black!70, fill=white, inner sep=0pt, minimum size=3.1mm, font=\figfontS},
  aptile/.style={draw=black!60, line width=0.5pt},
  apbrace/.style={decorate, decoration={brace, amplitude=1.2mm}, draw=black!60, line width=0.5pt},
}

\usepackage[english]{babel}
\newcommand{\sys}{AltoPIR\xspace}
\newcommand{\reins}{ReinsPIRe\xspace}
\newcommand{\inspire}{InsPIRe\xspace}
\newcommand{\valar}{Valar\xspace}
\newcommand{\ypir}{YPIR\xspace}
\newcommand{\ypirsp}{YPIR+SP\xspace}
\newcommand{\simplepir}{SimplePIR\xspace}
\newcommand{\sandwich}{SandwichPIR\xspace}
\newcommand{\port}{AMX-Lib\xspace}
\newcommand{\vnni}{\sys-VNNI\xspace}

\begin{document}
\title{\Large \bf High-Throughput Private Information Retrieval with on-CPU Accelerators}
\date{}
\author{
{\rm Gyuyeong Kim }\\ Yonsei University
}
\maketitle

\begin{abstract}
We present \sys, the first PIR system that exploits the capability of emerging on-CPU matrix accelerators such as Intel AMX. Unlike existing solutions, \sys achieves high-performance PIR through efficient batching without requiring discrete accelerators. However, the on-CPU accelerator has only a few small registers and shares its core with request handling. As a result, it runs far below its peak rate while its operands arrive from memory, and it remains idle while its core handles requests. To keep the accelerator busy, we design mechanisms that reduce both the memory traffic of a batch and the core cycles spent on each request. Our experimental results show that \sys outperforms existing CPU-based solutions by up to 50.1$\times$.
\end{abstract}

\section{Introduction}\label{introduction}
Private information retrieval (PIR) enables a client to retrieve a database record without revealing its choice to the server~\cite{chor1995pir,kushilevitz1997pir}.
Its representative applications are leaked-password checks~\cite{pantheon}, unsafe-link checks~\cite{checklist}, and privacy-preserving retrieval-augmented generation (RAG)~\cite{pirrag}.
Most PIR schemes require the server to read the entire database for each query~\cite{beimel2000pir}, making throughput critical to the cost of serving many concurrent users.
Such databases can be large (e.g., containing billions of leaked passwords~\cite{c3checks}), and the server typically keeps the entire database in memory.

A PIR system can improve throughput through \emph{cross-client batching}~\cite{luekspir,ypir}, which combines multiple clients' queries into one matrix multiplication over the database, called the \emph{scan}.
Recent schemes then compress the results into short responses with a second matrix product, called the \emph{packing}~\cite{ypir,inspire,reinspire}.
Packing multiplies a key from each client by a matrix that the server precomputes.
Unfortunately, batching shifts the bottleneck from memory to computation, and vector instruction sets (e.g., AVX-512) then limit the gain because they multiply only a few integer pairs at a time.

Some systems use GPUs~\cite{vipir,gpir,sandwichpir} or dedicated accelerators~\cite{ive} to overcome this.
Besides its general-purpose cores, a GPU has separate matrix units called tensor cores.
Tensor cores multiply 8-bit integers much faster than general-purpose cores multiply 32-bit integers.
Therefore, the latest GPU-based system~\cite{sandwichpir} splits each 4-byte (i.e., 32-bit) encrypted value into four bytes called \emph{limbs} and multiplies them as 8-bit integers on the tensor cores.
However, these systems require additional hardware, and a database larger than the memory of one device needs more devices.
In this context, we ask: \emph{Can a commodity CPU server achieve high throughput for batched PIR without a discrete accelerator?}

To answer this question affirmatively, we present \sys, the first PIR system that leverages the capability of on-CPU matrix accelerators.
An \emph{on-CPU matrix accelerator}, such as Intel AMX~\cite{intelamx}, is an emerging unit inside each core of recent CPUs.
It multiplies small matrices of low-precision numbers such as 8-bit integers, much like a small tensor core in GPUs.
This means we can do the limb split in CPUs.
It is an attractive platform for a PIR batch for three reasons.
First, its peak multiply-add rate is much higher than that of vector instruction sets.
Second, the built-in accelerator reads data from the CPU server's memory, which is far larger and cheaper per byte than the memory of a discrete accelerator.
Lastly, we can deploy \sys easily without additional hardware, as recent server CPUs already include it.

Unfortunately, two characteristics of an on-CPU accelerator make it hard to leverage for PIR.
First, the accelerator runs at its full rate only when its operands come from the L1 cache, but a PIR batch touches far more data than the caches can hold.
For example, Intel AMX holds only a few small operand registers, called \emph{tiles}.
A larger batch also leaves less cache space per query.
Furthermore, every result must be rebuilt exactly from its limb products, and the limb products are also too large to stay in the caches.
Second, the accelerator is often idle during a batch.
The CPU cores that feed the accelerator must also parse every request and assemble every response, and the accelerator waits while they do so.
As a result, the PIR throughput stays far below the peak rate.
For example, a straightforward port of a GPU-based design keeps the accelerator busy for only 22\% of a batch (\S\ref{sec:motiv-gap}).

To improve the utilization of the accelerator, \sys reduces the memory traffic of the scan and packing and frees the CPU cores from request handling with one design principle and four mechanisms.
Specifically, both products multiply a large shared operand (the database or a public matrix) by small per-client operands (encrypted queries or packing keys).
The principle is to keep the small operands in the caches and reuse each piece of the shared operand across all clients, reading it from memory only once per batch.
The \emph{query-blocked scan} (\S\ref{design:scan}) and \emph{column-group packing} (\S\ref{design:pack}) apply this principle.
Both recombine their limb products inside the core instead of writing them to memory.
For a large batch, the scan runs once per \emph{sub-batch} of 128 queries, which keeps it equally fast (\S\ref{design:placement}).
Lastly, the \emph{in-place engine} validates each request in its receive buffer and completes each response in one vector pass (\S\ref{design:engine}), freeing CPU cycles to feed the accelerator.

We implement a prototype of \sys on a commodity server and evaluate it against the state-of-the-art PIR systems~\cite{ypir,inspire,valar,reinspire}.
Experimental results show that \sys provides better throughput than the fastest of these systems by 24.2$\times$ to 50.1$\times$ on databases of 1\,GB to 64\,GB. \sys also keeps its advantage in three PIR applications, under random request arrivals, and alongside co-located jobs.

In summary, this work makes the following contributions:
\begin{itemize}[noitemsep,leftmargin=*]
\item We identify the matrix accelerators built into server CPUs as a platform for batched PIR without a discrete accelerator.
\item We design \sys, which places the scan and packing of a batch in the caches of a CPU core and answers requests in place with vector instructions.
\item We evaluate \sys against four state-of-the-art schemes and show that it outperforms the fastest of them.
\end{itemize}

\section{Background and Motivation \label{motivation}}

\subsection{Private Information Retrieval \label{sec:motiv-pir}}

\begin{figure}[t]
\centering
\begin{tikzpicture}[x=1cm,y=1cm]
\def\c{0.42}
\fill[black!5, rounded corners=2pt] (0.12,-0.62) rectangle (5.06,3.0);
\fill[black!5, rounded corners=2pt] (5.42,-0.62) rectangle (7.0,0.98);
\node[aplabel, anchor=north west] at (5.48,0.93) {Packing};
\node[aplabel] at (6.21,-0.09) {Packing keys\\$\times$ packing\\matrix};
\node[aplabel, anchor=north west] at (0.18,2.95) {Scan};
\foreach \r/\col in {0/ppBlue, 1/ppViolet, 2/ppGreen, 3/ppTeal} {
  \foreach \j in {0,...,5} {
    \draw[aptile, fill=\col] (2.4+\j*\c, 2.6-\r*\c-\c) rectangle ++(\c,\c);
  }
  \node[aplabel, anchor=east] at (2.32, 2.6-\r*\c-0.5*\c) {Record \r};
}
\node[aplabel, anchor=south] at (2.4+3*\c, 2.65) {Database};
\foreach \cl/\sel/\col [count=\i from 0] in {A/1/ppViolet, B/3/ppTeal, C/0/ppBlue} {
  \foreach \r in {0,...,3} {
    \ifnum\r=\sel
      \draw[aptile, fill=\col] (0.47+\r*\c, 0.67-\i*\c-\c) rectangle ++(\c,\c);
      \node[aplabel] at (0.47+\r*\c+0.5*\c, 0.67-\i*\c-0.5*\c) {\textbf{1}};
    \else
      \draw[aptile, fill=white] (0.47+\r*\c, 0.67-\i*\c-\c) rectangle ++(\c,\c);
      \node[aplabel] at (0.47+\r*\c+0.5*\c, 0.67-\i*\c-0.5*\c) {0};
    \fi
  }
  \node[aplabel, anchor=east] at (0.42, 0.67-\i*\c-0.5*\c) {\cl};
  \draw[aptile, fill=\col] (2.4, 0.67-\i*\c-\c) rectangle ++(6*\c,\c);
  \node[aplabel] at (2.4+3*\c, 0.67-\i*\c-0.5*\c) {E(Record \sel)};
}
\node[aplabel, anchor=north] at (0.47+2*\c, 0.67-3*\c-0.08) {Encrypted one-hot\\query vectors};
\node[aplabel, anchor=north] at (2.4+3*\c, 0.67-3*\c-0.08) {Scan output, one\\ciphertext per byte};
\draw[ppDarkRed, line width=0.9pt] (2.4+5*\c, 2.6-4*\c) rectangle ++(\c, 4*\c);
\draw[ppDarkRed, line width=0.9pt] (2.4+5*\c, 0.67-3*\c) rectangle ++(\c, 3*\c);
\draw[ppDarkRed, line width=0.6pt, densely dashed] (2.4+5.5*\c, 2.6-4*\c) -- (2.4+5.5*\c, 0.67);
\node[aplabel, anchor=west] at (5.08, 1.76) {Each byte read once\\serves all clients};
\draw[aparrow] (2.4+6*\c+0.06, 0.67-1.5*\c) -- (5.42, 0.67-1.5*\c);
\foreach \cl [count=\i from 0] in {A,B,C} {
  \node[apnode, minimum width=0.5cm, minimum height=0.36cm, inner sep=0.5pt] (r\i) at (7.6, 0.67-\i*\c-0.5*\c) {\cl};
  \draw[aparrow] (7.0, 0.67-\i*\c-0.5*\c) -- (r\i.west);
}
\node[aplabel, anchor=north] at (7.6, 0.67-3*\c-0.08) {Short\\responses};
\end{tikzpicture}
\caption{PIR queries answered with two matrix products.
}
\label{fig:pir}
\vspace{-4mm}
\end{figure}
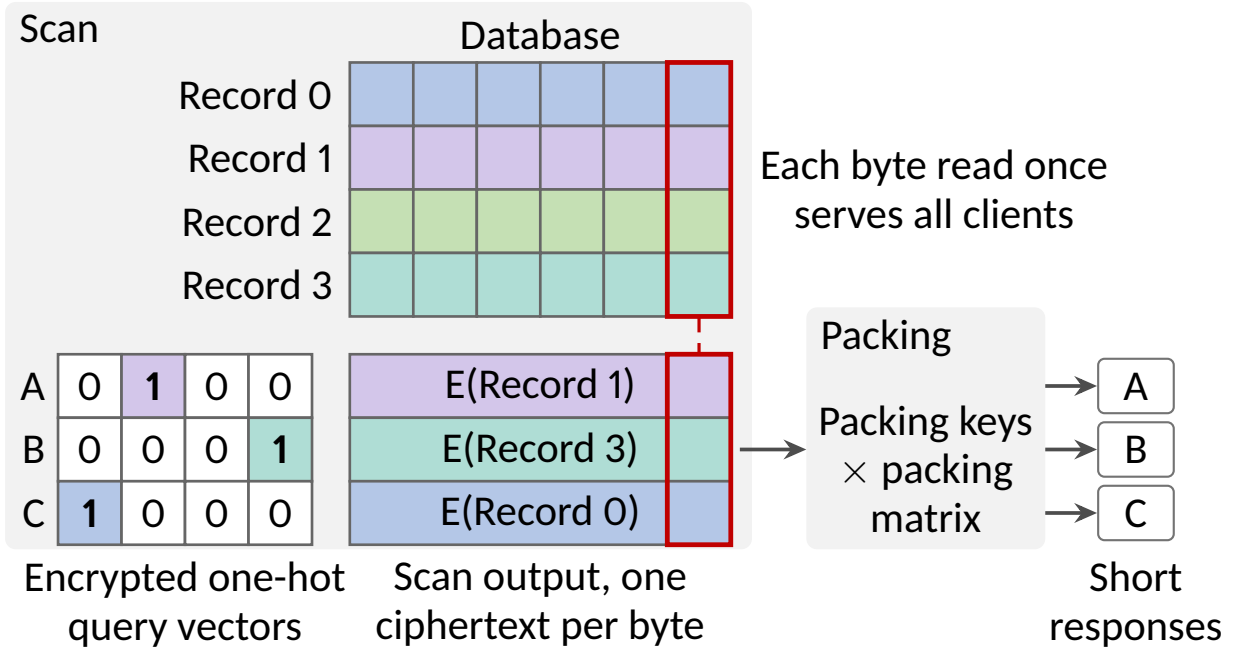

\noindent\textbf{PIR by example.}
A PIR protocol hides from the server which record a client fetches~\cite{chor1995pir}.
The schemes that we consider build this from a linearly homomorphic encryption scheme.
Such a scheme allows the server to multiply encrypted numbers by plain numbers and add them without decrypting~\cite{simplepir}.
Figure~\ref{fig:pir} shows the idea with a database of four records.
To fetch record~1, client~A sends the \emph{query vector} $(0,1,0,0)$ with every entry encrypted.
The server multiplies each record by the matching encrypted entry and adds up the results, which decrypt to record~1.
Because every entry looks random to the server, it must touch every record for every query.
This \emph{scan} yields one ciphertext per byte of the record.
Recent schemes are \emph{hintless}~\cite{hintlesspir,ypir,inspire,reinspire}.
Unlike hint-based schemes~\cite{simplepir,frodopir}, they need no digest of the database on the client.
Instead, the server \emph{packs} these ciphertexts into a few compact ones with a \emph{packing key} that the client sends.
The server still computes the hint but does not send it to the clients.
It uses the hint to derive a public matrix for packing, called the \emph{packing matrix}.

\begin{figure}[t]
\centering
\subfloat[Batching efficiency of \ypir~\cite{ypir}]{\includegraphics[width=0.48\linewidth]{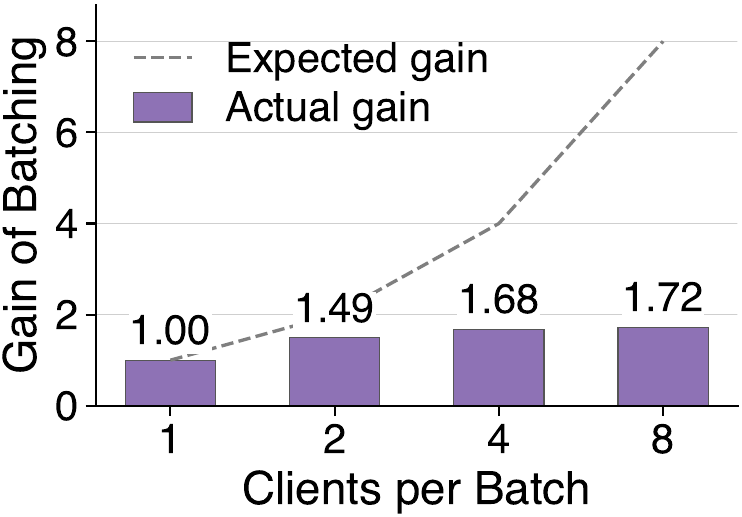}}\hfill
\subfloat[AMX and AVX-512 VNNI]{\includegraphics[width=0.48\linewidth]{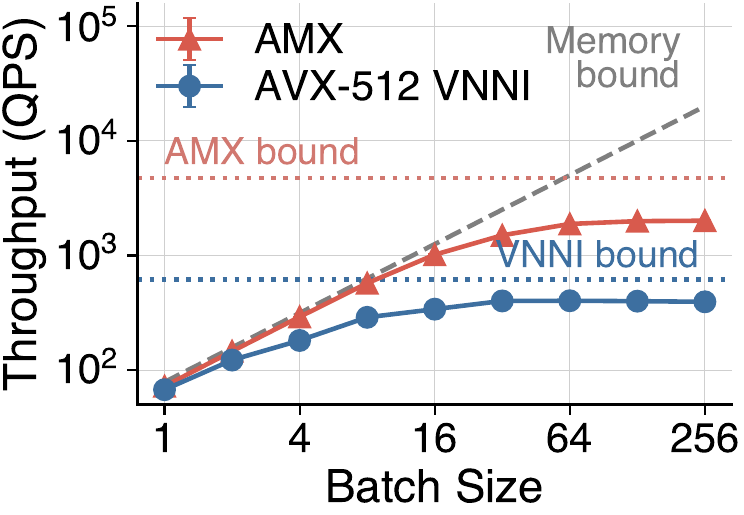}}
\caption{Existing CPU batching gains far less than expected, while AMX keeps gaining where VNNI stops.}
\label{fig:motiv-batch}
\vspace{-3mm}
\end{figure}

\noindent\textbf{Matrix computation.}
Both steps of the server are matrix products.
The scan multiplies the database, a matrix of bytes, by the query vectors.
Packing multiplies the packing key of the client by the packing matrix.
The scan reads the whole database.
In contrast, the packing matrix has one row per entry of the packing key and one column per byte of a record.
The size of the packing matrix depends only on the record size, not on the number of records.
A short per-request step, called the \emph{tail}, then combines the outputs of the scan and packing into the response.

The tensor cores of a GPU multiply only integers of up to 8 bits.
In exchange, they run many more multiplications per second than general-purpose cores.
To run the scan and packing on matrix units, a GPU-based system~\cite{sandwichpir} splits every 32-bit value $x$ into four bytes, called \emph{limbs}, with $x = x_0 + 2^{8}x_1 + 2^{16}x_2 + 2^{24}x_3$.
For example, the value $258=2+2^8$ has the limbs $(2,1,0,0)$.
Another GPU-based system~\cite{vipir} uses limbs only in the scan.
In the scan, each database entry is already a byte, and only the query value has four limbs.
Therefore, multiplying the two takes four limb products, which are added back with these weights modulo $2^{32}$.
In packing, both operands have 32-bit entries with four limbs each.
Therefore, every product takes $4\times4=16$ limb products, which are added back modulo a 32-bit prime.
The speed of matrix units more than makes up for these extra limb products.
\sys uses the same split on the on-CPU accelerator (\S\ref{sec:motiv-why}).

\subsection{Limitations of Existing Solutions \label{sec:motiv-vnni}}

A single query performs few multiply-adds per database byte; hence, its scan is limited by memory bandwidth.
\emph{Cross-client batching}~\cite{luekspir,ypir} stacks independent clients' query vectors into a matrix without changing any message (Figure~\ref{fig:pir}).
This allows one database read to serve many queries, reducing memory traffic per query while leaving the number of multiplications unchanged.
Therefore, batching moves the bottleneck from memory to computation and improves throughput only when the CPU has spare compute capacity.
Existing CPU systems quickly exhaust this capacity because they operate directly on 64- or 32-bit integers.
Specifically, one AVX-512 instruction multiplies only 8 or 16 such pairs.
With this, \ypir~\cite{ypir} reports at most 1.4$\times$ batching gains in its evaluation.
In our testbed server, \ypir processes eight clients only 1.72$\times$ faster than one at a time (Figure~\ref{fig:motiv-batch}~(a)).
These schemes also pack each client's response separately.
Some systems use GPUs (VIPIR~\cite{vipir}, \sandwich~\cite{sandwichpir}, GPIR~\cite{gpir}) or a dedicated accelerator (IVE~\cite{ive}) to increase compute capacity.
However, they require additional hardware, and each accelerator holds only as much of the database as fits in its memory.
For example, a GPU provides at most a few hundred gigabytes of memory, whereas a server CPU supports several terabytes of memory~\cite{intel4514yspec}.
Therefore, a larger database needs more accelerators.

\subsection{Why On-CPU Accelerators for PIR? \label{sec:motiv-why}}

\sys is a new PIR system that exploits the capability of on-CPU matrix accelerators.
An \emph{on-CPU matrix accelerator} is a unit built into a CPU core that multiplies small matrices held in dedicated registers.
Such accelerators are already built into recent CPUs.
Examples are Intel AMX~\cite{intelamx}, the Matrix-Multiply Assist of IBM Power10~\cite{power10mma}, and Arm SME~\cite{armsme}.
They also support 8-bit integer multiplication with 32-bit accumulation, which \sys uses to multiply database bytes by the byte limbs of encrypted queries.
\sys targets Intel AMX, which has eight \emph{tile} registers of 1\,KB, each holding a matrix of 16 rows of 64 bytes.
One tile instruction multiplies two such matrices and adds the $16\times16$ result to a third tile.
Per cycle, the peak 8-bit multiply-add rate of AMX is higher than that of the AVX-512 VNNI vector instructions by 8$\times$.
Figure~\ref{fig:motiv-batch}~(b) runs \sys on 16 CPU cores with a 1\,GB database of 32\,KB records, once with AMX and once with VNNI.
We can observe that, with VNNI, the throughput gains only 6.0$\times$ from batching.
In contrast, \sys with AMX gains up to 27.9$\times$.

\begin{figure}[t]
\centering
\subfloat[One batch of \port]{\includegraphics[width=0.48\linewidth]{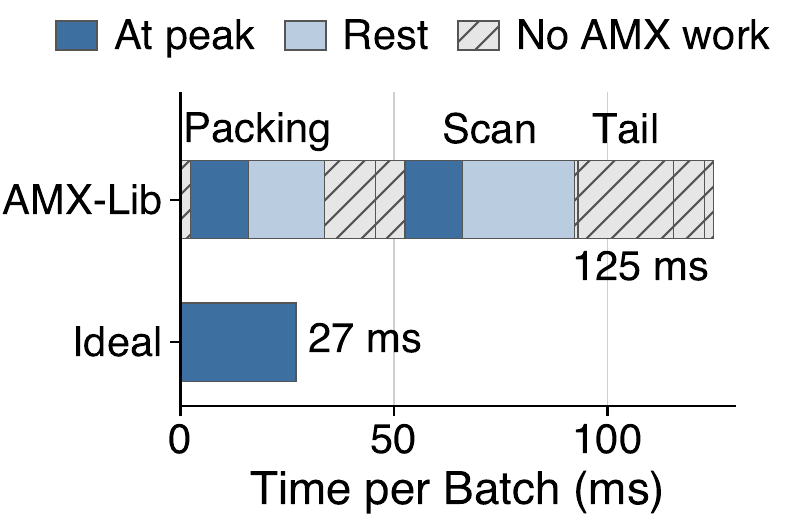}}\hfill
\subfloat[Rate of tile multiplications]{\includegraphics[width=0.48\linewidth]{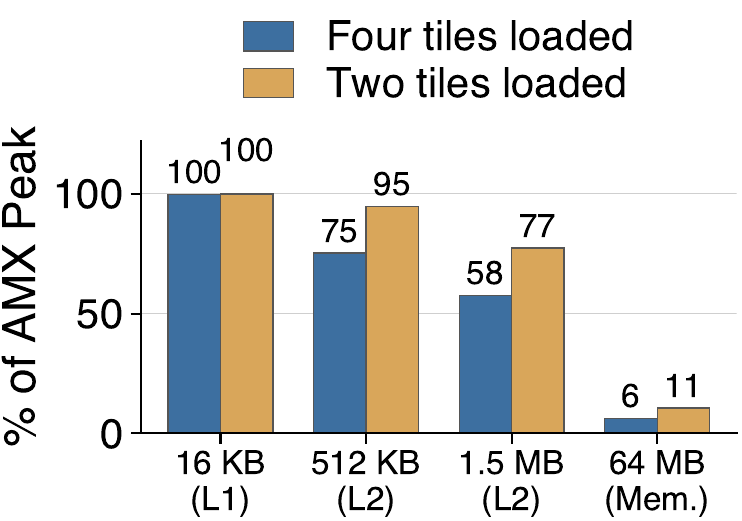}}
\caption{\port leaves the accelerator idle for most of a batch. In (b), operand tiles from the L2 cache or memory keep the accelerator below its peak rate.}
\label{fig:motiv-port}
\vspace{-3mm}
\end{figure}

\subsection{A Straightforward Design Is Not Enough \label{sec:motiv-gap}}

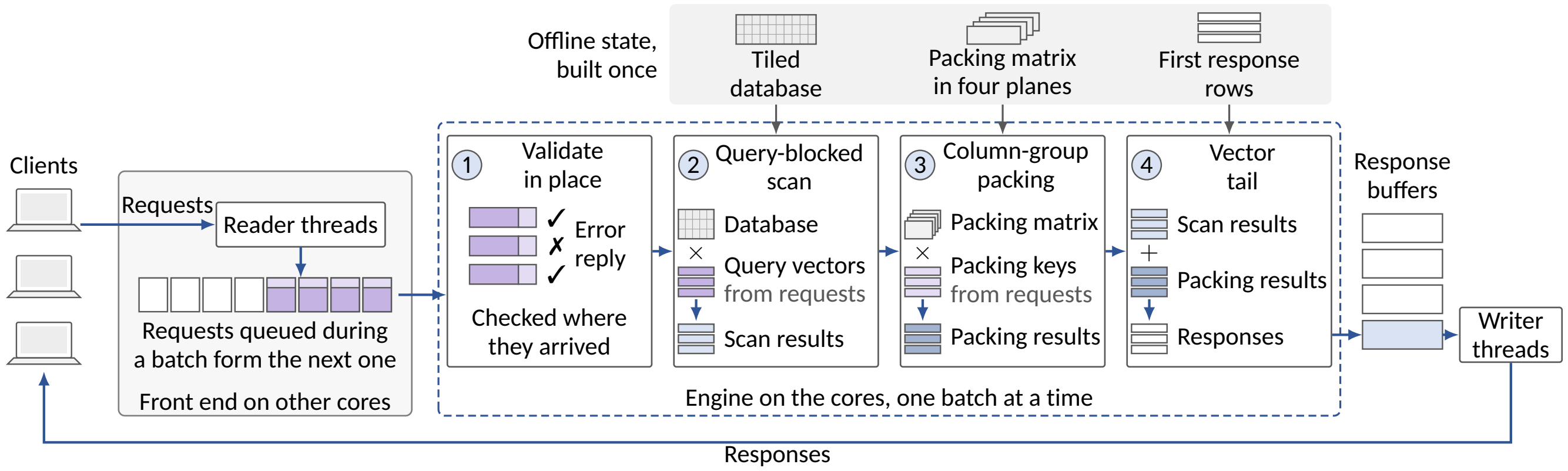
\begin{figure*}[t]
\centering
\begin{tikzpicture}[x=1cm,y=1cm]
\newcommand{\bstack}[3]{\foreach \r in {0,1,2} \draw[dtile, fill=#3] (#1,#2+\r*0.12) rectangle ++(0.4,0.09);}
\foreach \y in {1.95,1.2,0.45} {
  \draw[dtile, fill=white, rounded corners=0.6pt] (0.3,\y) rectangle ++(0.62,0.4);
  \fill[apStoneL] (0.35,\y+0.05) rectangle ++(0.52,0.3);
  \draw[dtile, fill=apStoneL!70] (0.2,\y-0.07) -- (1.02,\y-0.07) -- (0.95,\y) -- (0.27,\y) -- cycle;
}
\node[dlab] at (0.61,2.62) {Clients};
\draw[draw=apStone, line width=0.5pt, fill=apStoneL!45, rounded corners=2pt] (1.45,-0.2) rectangle (4.75,2.55);
\node[dlab, anchor=south] at (3.1,-0.17) {Front end on other cores};
\node[dbox, minimum width=1.9cm, minimum height=0.5cm] (rd) at (3.5,1.95) {Reader threads};
\draw[dgo] (1.02,1.95) -- (rd.west);
\node[dlab, anchor=south] at (2.0,2.0) {Requests};
\foreach \k in {0,...,7} {
  \draw[dtile, fill=white] (1.68+\k*0.36,0.97) rectangle ++(0.32,0.38);
  \ifnum\k>3
    \fill[apLimb!45] (1.68+\k*0.36,0.97) rectangle ++(0.32,0.27);
    \fill[apLimbL] (1.68+\k*0.36,1.24) rectangle ++(0.32,0.11);
    \draw[dtile] (1.68+\k*0.36,0.97) rectangle ++(0.32,0.38);
    \draw[dtile] (1.68+\k*0.36,1.24) -- ++(0.32,0);
  \fi
}
\draw[dgo] (3.5,1.7) -- (3.5,1.35);
\node[dlab] at (3.1,0.6) {Requests queued during\\a batch form the next one};
\fill[apStoneL!70, rounded corners=2pt] (7.65,3.3) rectangle (15.1,4.43);
\node[dlab, anchor=east, align=right] at (7.55,3.86) {Offline state,\\built once};
\foreach \j in {0,...,9} \foreach \r in {0,...,2} \draw[dgrid, fill=apStoneL] (8.4+\j*0.09,3.98+\r*0.11) rectangle ++(0.09,0.11);
\draw[dtile] (8.4,3.98) rectangle ++(0.9,0.33);
\node[dlab] at (8.85,3.65) {Tiled\\database};
\foreach \k in {3,2,1,0} \draw[dtile, fill=apStoneL!80] (11.0+\k*0.07,3.98+\k*0.05) rectangle ++(0.6,0.2);
\node[dlab] at (11.4,3.65) {Packing matrix\\in four planes};
\foreach \k in {0,...,2} \draw[dtile, fill=white] (13.6,4.0+\k*0.12) rectangle ++(0.7,0.09);
\node[dlab] at (13.95,3.65) {First response\\rows};
\draw[draw=apPetrol, line width=0.6pt, dash pattern=on 2.5pt off 1.4pt, rounded corners=2pt] (5.05,-0.2) rectangle (15.2,3.1);
\node[dlab, anchor=south] at (10.125,-0.17) {Engine on the cores, one batch at a time};
\foreach \c/\n/\t [count=\i] in {6.3/s1/{Validate\\in place}, 8.85/s2/{Query-blocked\\scan}, 11.4/s3/{Column-group\\packing}, 13.95/s4/{Vector\\tail}} {
  \node[dbox, fill=white, minimum width=2.3cm, minimum height=2.6cm] (\n) at (\c,1.65) {};
  \node[dlab] at (\c+0.15,2.6) {\t};
  \node[circle, draw=apEdge, line width=0.45pt, fill=apPetrolL, inner sep=0pt, minimum size=3.3mm, font=\figfontS] at (\c-0.93,2.62) {\i};
}
\foreach \a/\b in {s1/s2, s2/s3, s3/s4} \draw[dgo] (\a.east) -- (\b.west);
\draw[dgo] (4.6,1.16) -- (s1.west |- 0,1.16);
\foreach \y/\m in {1.92/{\ding{51}}, 1.6/{\ding{55}}, 1.28/{\ding{51}}} {
  \draw[dtile, fill=apLimb!45] (5.4,\y) rectangle ++(0.55,0.22);
  \draw[dtile, fill=apLimbL] (5.95,\y) rectangle ++(0.2,0.22);
  \node[dlab] at (6.38,\y+0.11) {\m};
}
\node[dlab, anchor=west] at (6.55,1.71) {Error\\reply};
\node[dlab] at (6.3,0.72) {Checked where\\they arrived};
\foreach \c/\ta/\tb/\op/\tc in {8.85/Database/{Query vectors\\\textcolor{apEdge}{from requests}}/$\times$/Scan results, 11.4/Packing matrix/{Packing keys\\\textcolor{apEdge}{from requests}}/$\times$/Packing results, 13.95/Scan results/Packing results/$+$/Responses} {
  \node[dlab, anchor=west] at (\c-0.62,1.96) {\ta};
  \node[dlab] at (\c-0.9,1.63) {\op};
  \node[dlab, anchor=west, align=left] at (\c-0.62,1.31) {\tb};
  \draw[dgo] (\c-0.9,1.11) -- (\c-0.9,0.86);
  \node[dlab, anchor=west] at (\c-0.62,0.67) {\tc};
}
\foreach \j in {0,...,4} \foreach \r in {0,...,2} \draw[dgrid, fill=apStoneL] (7.75+\j*0.08,1.79+\r*0.11) rectangle ++(0.08,0.11);
\draw[dtile] (7.75,1.79) rectangle ++(0.4,0.33);
\bstack{7.75}{1.14}{apLimb!45}
\bstack{7.75}{0.5}{apPetrolL}
\foreach \k in {3,2,1,0} \draw[dtile, fill=apStoneL!80] (10.3+\k*0.03,1.79+\k*0.04) rectangle ++(0.31,0.2);
\bstack{10.3}{1.14}{apLimbL}
\bstack{10.3}{0.5}{apPetrol!45}
\bstack{12.85}{1.79}{apPetrolL}
\bstack{12.85}{1.14}{apPetrol!45}
\bstack{12.85}{0.5}{white}
\foreach \x/\n in {8.85/s2, 11.4/s3, 13.95/s4} \draw[dplain] (\x,3.3) -- (\n.north);
\node[dlab] at (15.9,2.5) {Response\\buffers};
\draw[dtile, fill=apPetrolL] (15.45,0.55) rectangle ++(0.9,0.32);
\foreach \k in {1,...,3} \draw[dtile, fill=white] (15.45,0.55+\k*0.4) rectangle ++(0.9,0.32);
\draw[dgo] (s4.east |- 0,0.71) -- (15.45,0.71);
\node[dbox, minimum width=1.15cm, minimum height=0.55cm] (wr) at (17.12,0.71) {Writer\\threads};
\draw[dgo] (16.35,0.71) -- (wr.west);
\draw[dgo] (wr.south) -- (17.12,-0.5) -- node[dlab, below, pos=0.5] {Responses} (0.61,-0.5) -- (0.61,0.3);
\end{tikzpicture}
\caption{\sys architecture. The engine answers each batch of queued requests in four steps (\S\ref{design:overview}). %
}
\label{fig:overview}
\end{figure*}

An on-CPU matrix accelerator works like a small tensor core.
Therefore, a straightforward solution would port the GPU data flow and run each of the scan and packing with the faster of the vendor libraries oneDNN~\cite{onednn} and oneMKL~\cite{onemkl}.
\port is our optimized CPU port of the data flow of \sandwich~\cite{sandwichpir}.
However, Figure~\ref{fig:motiv-port}~(a) shows that \port keeps the accelerator busy for only 22\% of a batch of 128 queries on a 1\,GB database.

There are three reasons behind this low utilization.
First, the accelerator holds only eight tile registers, unlike a GPU, which keeps blocks of operands in a large register file and shared memory.
Operand tiles from the L2 cache allow only 58\% to 75\% of its peak rate, and those from memory only 6\%, as Figure~\ref{fig:motiv-port}~(b) shows.
Second, the libraries do not suit the shapes of the scan and packing in a PIR batch.
In each product, the database or the packing matrix is a large shared operand, and the query limbs or the packing keys are small per-client operands.
The scan runs at 34\% of the peak rate and reads the database from memory 2.6 times instead of once.
Following the GPU data flow, packing writes all 16 limb products of every output to memory.
Third, the accelerator shares its cores with request handling, which a GPU leaves to the host CPU.
Request handling and the per-request tail take 37\,ms, more than the scan and packing would need at the peak rate.
These reasons imply that an on-CPU accelerator speeds up PIR only when the system is designed around how it differs from a GPU.

\begin{table}[t]
\centering
\caption{Comparison to existing approaches.\label{tab:comparison}}
\footnotesize
\setlength{\tabcolsep}{0.9pt}
\begin{tabular}{@{}l c c c c@{}}
\toprule
 & \shortstack{\ypir~\cite{ypir},\\\inspire~\cite{inspire},\\\reins~\cite{reinspire}}
 & \shortstack{\sandwich~\cite{sandwichpir},\\GPIR~\cite{gpir},\\VIPIR~\cite{vipir}}
 & \shortstack{IVE\\\cite{ive}}
 & \sys \\
\midrule
Platform & CPU & GPU & ASIC & CPU \\
\midrule
High throughput & \xmark & \cmark & \cmark & \cmark \\
Readily deployable & \cmark & \cmark & \xmark & \cmark \\
Commodity CPU server & \cmark & \xmark & \xmark & \cmark \\
Large database support & \cmark & \xmark & \xmark & \cmark \\
\bottomrule
\end{tabular}
\vspace{-3mm}
\end{table}

\noindent\textbf{Comparison to existing work.}
Table~\ref{tab:comparison} compares \sys with existing approaches.
CPU-based schemes~\cite{ypir,inspire,reinspire} run on commodity servers and keep the database in main memory, but only \ypir batches queries, which raises its throughput only a little (\S\ref{sec:motiv-vnni}).
GPU-based systems~\cite{sandwichpir,gpir,vipir} achieve high throughput with GPUs.
However, every server then needs a GPU, and a database larger than the memory of one GPU needs more GPUs.
IVE~\cite{ive} needs a custom chip that servers do not have today, and the same memory limit applies.
\sys meets all four requirements.
It runs the scan and packing on the accelerator inside each CPU core and serves the database from the main memory of the server.

\section{\sys Design \label{design}}

\subsection{Overview \label{design:overview}}
\sys answers the queries of many independent clients together on CPU matrix accelerators.
It follows the hintless scheme of \sandwich~\cite{sandwichpir}, which defines what the server computes for each query.
The design of \sys decides how a CPU computes these responses quickly.
Our design principle is to read each shared operand once per batch and to reuse the per-client operands from the caches.
Figure~\ref{fig:overview} shows the architecture of \sys.

\noindent\textbf{Offline state.}
When the database is loaded, \sys builds the three pieces of offline state.
First, it stores the database in \emph{tile order}, in which every tile that the accelerator loads lies contiguously in memory.
Second, it precomputes the packing matrix and splits its 32-bit entries into four byte planes.
The first plane holds the lowest byte of every entry, and the second plane holds the next byte.
All four planes are in tile order.
Third, it precomputes the first row of every response ciphertext, which is the same for all clients.
All three depend only on the database and the public parameters.
Because they are built once, a batch neither changes any layout nor recomputes anything that does not depend on the clients.

\noindent\textbf{A request workflow.}
We now describe how a request is handled by \sys.
A client sends one request that holds its encrypted query vector and two keys.
On the left of Figure~\ref{fig:overview}, reader threads append arriving requests to a bounded queue, and a request that arrives when it is full receives a busy reply.
When the engine finishes a batch, it takes every waiting request as the next batch, up to a maximum batch size.
It waits for neither a fixed size nor a timer.
At a low arrival rate, a timer would add its wait to every query.
At a high rate, the next batch fills during the current one even without a timer.
In this way, batches stay small at a low rate and grow toward the maximum as the rate grows.

The engine answers a batch in four steps.
\textcircled{\footnotesize 1}~It checks every request frame where the frame arrived, without copying or decoding it, and rejects malformed frames (\S\ref{design:engine}).
\textcircled{\footnotesize 2}~It runs the scan (\S\ref{design:scan} and \S\ref{design:placement}).
The scan reads the database from memory once and multiplies it by the query limbs held in the L2 cache.
It recombines the limb sums before they leave the core.
\textcircled{\footnotesize 3}~It runs packing (\S\ref{design:pack}).
Packing reads the packing matrix from memory once and multiplies it by the packing keys held in the L3 cache.
It recombines the limb sums of each output while they are still in the L2 cache.
\textcircled{\footnotesize 4}~It runs a tail with vector instructions for each client (\S\ref{design:engine}).
The tail reads the outputs of the scan and packing and writes the response bytes into one of several rotating buffers.
Writer threads send each response from that buffer, and the client decrypts it.

\subsection{Query-Blocked Scan \label{design:scan}}

The scan computes one output for every query and every byte position of a record.
Each output adds up, over all records, the byte of the record at that position times the query entry of that record.
Because the accelerator multiplies only bytes, every query entry is split into its four limbs (\S\ref{sec:motiv-pir}).
The scan then keeps four separate sums per output, one per limb, called the \emph{limb sums}.
Adding them back with their weights gives the output, $s_0 + 2^{8}s_1 + 2^{16}s_2 + 2^{24}s_3$, a step that we call \emph{recombining}.
For example, the query entry 258 has the limbs $(2,1,0,0)$.
Multiplied by the database byte 3, they give $(6,3,0,0)$, which recombine into $6 + 2^{8}\cdot3 = 774 = 258\cdot3$.

The scan should keep the accelerator busy although the database is far larger than the caches.
Because operand tiles from memory allow only 6\% of the peak rate (\S\ref{sec:motiv-gap}), each database tile should be read from memory once and meet the query limbs of the whole batch while it is in the core.
However, these query limbs do not fit in the L2 cache.
For example, 128 queries on a 1\,GB database of 32\,KB records have 16\,MB of query limbs, eight times a 2\,MB L2 cache.
Our observation is that each record meets only its own entry of every query.
For example, record~2 in Figure~\ref{fig:pir} is multiplied only by entry~2 of each query vector.
Therefore, the scan cuts the records into blocks and keeps in the L2 cache only the query limbs of one block.
It reads the block of the database from memory once and multiplies it by these limbs.
At the end of the block, it recombines the limb sums before they leave the core.
We call this a \emph{query-blocked scan}.
Keeping a part of the database in the core instead would read all query limbs from memory once per part, which is far more traffic than reading the database once.

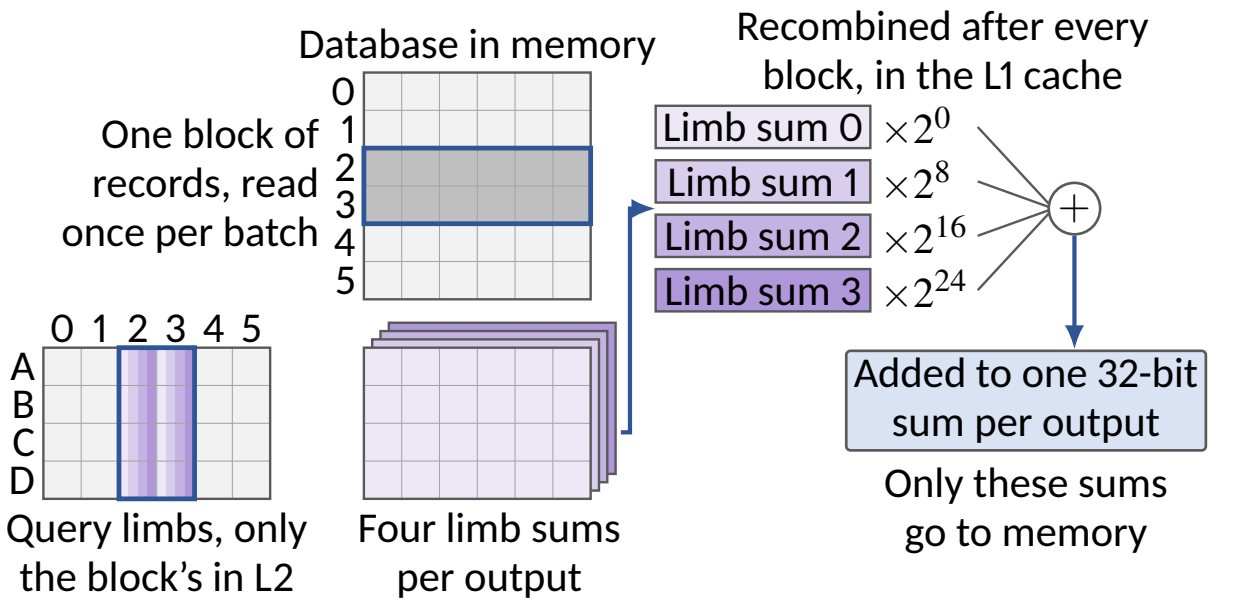
\begin{figure}[t]
\centering
\begin{tikzpicture}[x=1cm,y=1cm]
\def\c{0.25}
\fill[apStoneL!70] (0.50,0.30) rectangle ({0.50+6*\c},{0.30+4*\c});
\foreach \r in {2,3} \foreach \l/\s in {0/14,1/30,2/46,3/62}
  \fill[apLimb!\s] ({0.50+\r*\c+\l*\c/4},0.30) rectangle ++({\c/4},{4*\c});
\foreach \r in {1,...,5} \draw[dgrid] ({0.50+\r*\c},0.30) -- ++(0,{4*\c});
\foreach \i in {1,2,3} \draw[dgrid] (0.50,{0.30+\i*\c}) -- ++({6*\c},0);
\draw[dtile] (0.50,0.30) rectangle ({0.50+6*\c},{0.30+4*\c});
\draw[apPetrol, line width=0.9pt] ({0.50+2*\c},0.30) rectangle ({0.50+4*\c},{0.30+4*\c});
\foreach \q [count=\i from 0] in {A,B,C,D} \node[dlab, anchor=east] at (0.48,{0.30+4*\c-\i*\c-0.5*\c}) {\q};
\foreach \r in {0,...,5} \node[dlab] at ({0.50+\r*\c+0.5*\c},{0.30+4*\c+0.13}) {\r};
\node[dlab, anchor=north] at (1.25,0.24) {Query limbs, only\\the block's in L2};
\def\dx{2.62} \def\dt{3.12}
\fill[apStoneL!70] (\dx,{\dt-6*\c}) rectangle ({\dx+6*\c},\dt);
\fill[apStone!50] (\dx,{\dt-4*\c}) rectangle ({\dx+6*\c},{\dt-2*\c});
\foreach \j in {1,...,5} \draw[dgrid] ({\dx+\j*\c},{\dt-6*\c}) -- ++(0,{6*\c});
\foreach \r in {1,...,5} \draw[dgrid] (\dx,{\dt-\r*\c}) -- ++({6*\c},0);
\draw[dtile] (\dx,{\dt-6*\c}) rectangle ({\dx+6*\c},\dt);
\draw[apPetrol, line width=0.9pt] (\dx,{\dt-4*\c}) rectangle ({\dx+6*\c},{\dt-2*\c});
\foreach \r in {0,...,5} \node[dlab, anchor=east] at ({\dx-0.02},{\dt-\r*\c-0.5*\c}) {\r};
\node[dlab] at ({\dx+3*\c},{\dt+0.17}) {Database in memory};
\node[dlab, anchor=east, align=right] at ({\dx-0.26},{\dt-3*\c}) {One block of\\records, read\\once per batch};
\foreach \k/\s in {3/62,2/46,1/30} \draw[dtile, fill=apLimb!\s] ({\dx+\k*0.055},{0.30+\k*0.055}) rectangle ++({6*\c},{4*\c});
\fill[apLimb!14] (\dx,0.30) rectangle ({\dx+6*\c},{0.30+4*\c});
\foreach \j in {1,...,5} \draw[dgrid] ({\dx+\j*\c},0.30) -- ++(0,{4*\c});
\foreach \i in {1,2,3} \draw[dgrid] (\dx,{0.30+\i*\c}) -- ++({6*\c},0);
\draw[dtile] (\dx,0.30) rectangle ({\dx+6*\c},{0.30+4*\c});
\node[dlab, anchor=north] at (3.45,0.24) {Four limb sums\\per output};
\node[dlab] at (6.45,3.25) {Recombined after every\\block, in the L1 cache};
\foreach \l/\s/\w in {0/14/{$\times 2^{0}$},1/30/{$\times 2^{8}$},2/46/{$\times 2^{16}$},3/62/{$\times 2^{24}$}} {
  \draw[dtile, fill=apLimb!\s] (4.55,{2.62-\l*0.36}) rectangle ++(1.42,0.28);
  \node[dlab] at (5.26,{2.76-\l*0.36}) {Limb sum \l};
  \node[dlab, anchor=west] at (6.0,{2.76-\l*0.36}) {\w};
  \draw[apEdge, line width=0.45pt] (6.68,{2.76-\l*0.36}) -- (7.18,2.22);
}
\node[circle, draw=apEdge, line width=0.5pt, fill=white, inner sep=0pt, minimum size=3.4mm, font=\figfontS] (sum) at (7.32,2.22) {$+$};
\draw[dgo] ({\dx+6*\c+0.2},0.74) -| (4.38,2.22) -- (4.55,2.22);
\node[dbox, fill=apPetrolL, minimum width=2.3cm] (acc) at (7.0,0.95) {Added to one 32-bit\\sum per output};
\draw[dgo] (sum.south) -- (sum.south |- acc.north);
\node[dlab] at (7.0,0.22) {Only these sums\\go to memory};
\end{tikzpicture}%
\caption{Query-blocked scan of one block (records 2 and 3). Only the query limbs of these records stay in the L2 cache. The block of the database is read from memory once. The four limb sums of each output are recombined after the block, and only one 32-bit sum per output reaches memory.}
\label{fig:scan}
\end{figure}

\noindent\textbf{Recombining after every block.}
Figure~\ref{fig:scan} follows one block.
The accelerator holds two database tiles and two query tiles and multiplies every pair of them, each into its own accumulator tile.
As a result, each loaded tile serves two products.
At the end of the block, the engine copies the limb sums into a small buffer in the L1 cache.
It weights them and adds them to a running 32-bit sum per output.
These sums overflow 32 bits but stay exact.
This is because the scheme computes the scan modulo $2^{32}$, and a 32-bit integer that wraps around keeps exactly this remainder.
For example, a sum that reaches $2^{32}+5$ keeps 5, which is its correct value modulo $2^{32}$.
Therefore, the records can be cut into blocks anywhere, and only the 32-bit sums, a quarter of the size of the limb sums, leave the core.
After the last block, one rounding step switches each sum from modulus $2^{32}$ to the prime modulus of packing.

\noindent\textbf{Choosing the block length.}
A longer block recombines less often.
However, all of its query limbs must stay in the L2 cache next to the database tiles in flight.
The L2 cache is the largest cache that a core does not share.
Therefore, \sys sizes each block so that its query limbs take half of the L2 cache and the database tiles an eighth, and neither evicts the other.
For example, a core with an L2 cache of 2\,MB gives 1\,MB to the query limbs.
With 128 queries of four limbs each, every record needs 512 bytes of query limbs, and a block holds 2,048 records.
A 1\,GB database of 32\,KB records then takes 16 blocks, and each block multiplies 64\,MB of the database by the same 1\,MB of query limbs.

\subsection{Column-Group Packing \label{design:pack}}

Packing multiplies the packing key of each client, a vector of 8,192 values, by the packing matrix, which has one column per byte of a record.
Each output is the sum, over the 8,192 rows, of a key value times a matrix value, modulo a 32-bit prime.
Unlike in the scan, both operands hold 32-bit values with four limbs each.
Therefore, each output has $4\times4=16$ limb sums, one for each pair of a key limb and a matrix limb.
Following the GPU data flow, a naive design writes the limb sums of all outputs to memory and recombines them in a separate pass (\S\ref{sec:motiv-gap}).
This moves hundreds of megabytes to and from memory while the accelerator waits.
Our observation is that each output uses only one column of the packing matrix.
Therefore, \sys gives each core its own groups of 32 columns.
A core computes all 16 limb sums of the outputs of one group and recombines them while they are still in its L2 cache.
No limb sum ever reaches memory.
We call this \emph{column-group packing}.
A split along the rows would instead leave a partial sum of each output on several cores, which would have to combine them through memory.

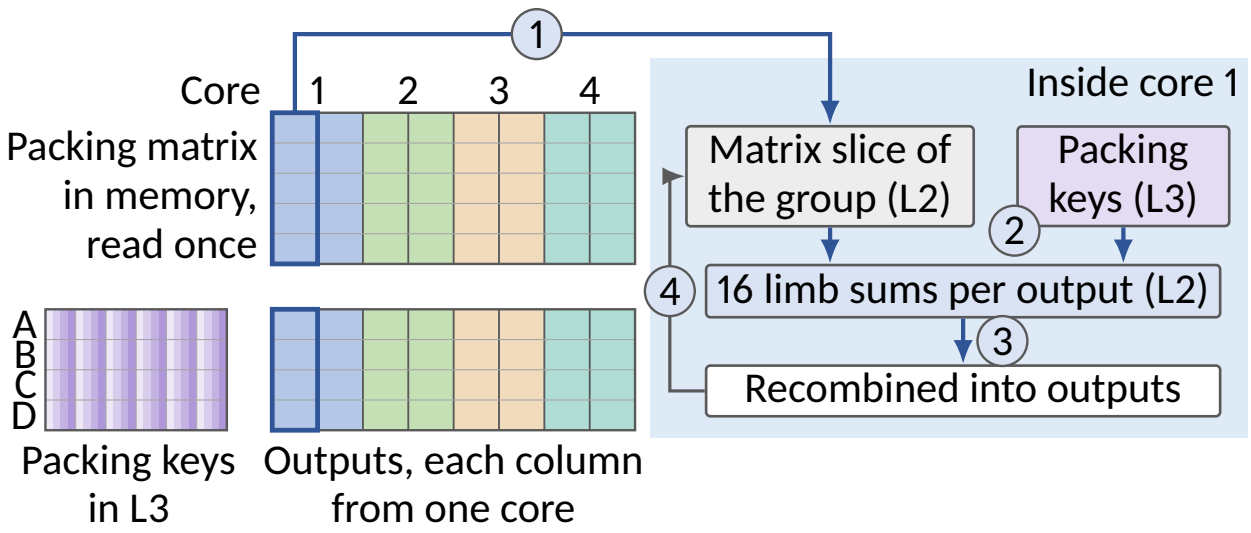
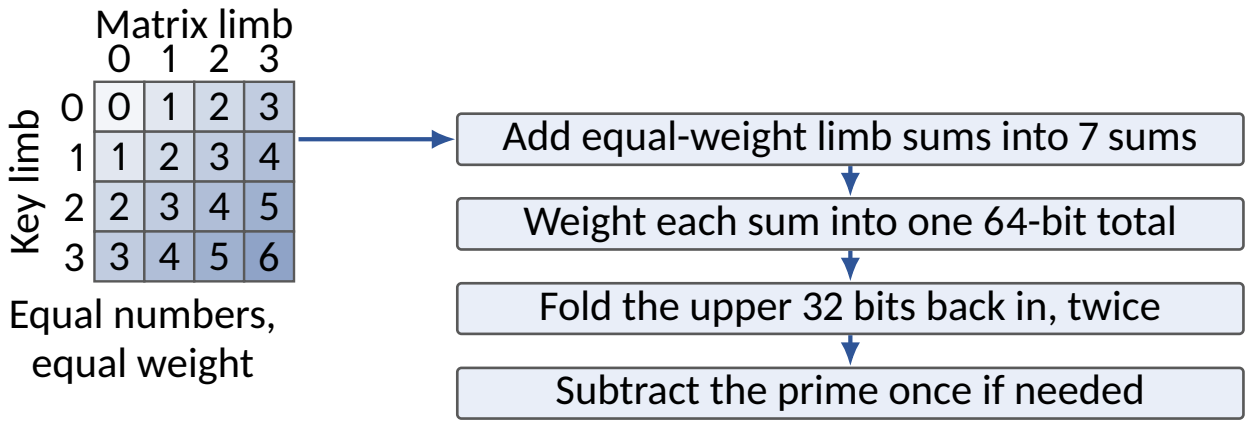
\begin{figure}[t]
\centering
\subfloat[Column groups split across cores]{\begin{tikzpicture}[x=1cm,y=1cm,
  pnum/.style={circle, draw=apEdge, line width=0.45pt, fill=apPetrolL, inner sep=0pt, minimum size=3.3mm, font=\figfontS}]
\def\r{0.2} \def\g{0.30} \def\px{1.95} \def\py{1.40}
\foreach \k in {0,...,5} \foreach \l/\s in {0/14,1/30,2/46,3/62}
  \fill[apLimb!\s] ({0.45+\k*0.2+\l*0.05},0.30) rectangle ++(0.05,{4*\r});
\foreach \k in {1,...,5} \draw[dgrid] ({0.45+\k*0.2},0.30) -- ++(0,{4*\r});
\foreach \i in {1,2,3} \draw[dgrid] (0.45,{0.30+\i*\r}) -- ++(1.2,0);
\draw[dtile] (0.45,0.30) rectangle (1.65,{0.30+4*\r});
\foreach \q [count=\i from 0] in {A,B,C,D} \node[dlab, anchor=east] at (0.43,{0.30+4*\r-\i*\r-0.5*\r}) {\q};
\node[dlab, anchor=north] at (1.0,0.24) {Packing keys\\in L3};
\foreach \gi/\col in {0/ppBlue,1/ppBlue,2/ppGreen,3/ppGreen,4/kbGold!40,5/kbGold!40,6/ppTeal,7/ppTeal} {
  \fill[\col] ({\px+\gi*\g},\py) rectangle ++(\g,{5*\r});
  \fill[\col] ({\px+\gi*\g},0.30) rectangle ++(\g,{4*\r});
}
\foreach \gi in {1,...,7} {
  \draw[apEdge, line width=0.35pt] ({\px+\gi*\g},\py) -- ++(0,{5*\r});
  \draw[apEdge, line width=0.35pt] ({\px+\gi*\g},0.30) -- ++(0,{4*\r});
}
\foreach \i in {1,...,4} \draw[dgrid] (\px,{\py+\i*\r}) -- ++({8*\g},0);
\foreach \i in {1,2,3} \draw[dgrid] (\px,{0.30+\i*\r}) -- ++({8*\g},0);
\draw[dtile] (\px,\py) rectangle ({\px+8*\g},{\py+5*\r});
\draw[dtile] (\px,0.30) rectangle ({\px+8*\g},{0.30+4*\r});
\draw[apPetrol, line width=0.9pt] (\px,\py) rectangle ({\px+\g},{\py+5*\r});
\draw[apPetrol, line width=0.9pt] (\px,0.30) rectangle ({\px+\g},{0.30+4*\r});
\node[dlab, anchor=east] at ({\px-0.04},{\py+5*\r+0.15}) {Core};
\foreach \c in {1,...,4} \node[dlab] at ({\px+(2*\c-1)*\g},{\py+5*\r+0.15}) {\c};
\node[dlab, anchor=east, align=right] at ({\px-0.06},{\py+2.5*\r-0.04}) {Packing matrix\\in memory,\\read once};
\node[dlab, anchor=north] at ({\px+4*\g},0.24) {Outputs, each column\\from one core};
\fill[apCore] (4.45,0.25) rectangle (8.40,2.76);
\node[dlab, anchor=north east] at (8.38,2.74) {Inside core 1};
\node[dbox, fill=apStoneL, minimum width=1.9cm] (sl) at (5.64,1.98) {Matrix slice of\\the group (L2)};
\node[dbox, fill=apLimbL, minimum width=1.4cm] (ky) at (7.58,1.98) {Packing\\keys (L3)};
\node[dbox, fill=apPetrolL, minimum width=3.4cm] (ls) at (6.52,1.22) {16 limb sums per output (L2)};
\node[dbox, fill=white, minimum width=3.4cm] (rc) at (6.52,0.56) {Recombined into outputs};
\draw[dgo] ({\px+0.5*\g},{\py+5*\r}) -- ({\px+0.5*\g},2.92) -| (sl.north);
\draw[dgo] (sl.south) -- (sl.south |- ls.north);
\draw[dgo] (ky.south) -- (ky.south |- ls.north);
\draw[dgo] (ls) -- (rc);
\draw[dplain] (rc.west) -- (4.58,0.56) |- (sl.west);
\node[pnum] at (3.70,2.92) {1};
\node[pnum] at (6.86,1.62) {2};
\node[pnum] at (6.78,0.89) {3};
\node[pnum] at (4.58,1.22) {4};
\end{tikzpicture}%
}\\
\subfloat[Recombining one output exactly]{\begin{tikzpicture}[x=1cm,y=1cm]
\def\c{0.33}
\foreach \i in {0,...,3} {
  \node[dlab, anchor=east] at (0.62,-0.75-\i*\c) {\i};
  \node[dlab, anchor=south] at (0.65+\i*\c+0.5*\c,-0.55) {\i};
  \foreach \j in {0,...,3} {
    \pgfmathtruncatemacro{\s}{\i+\j}
    \pgfmathtruncatemacro{\shade}{6+8*\s}
    \draw[dtile, fill=apPetrol!\shade] (0.65+\j*\c,-0.58-\i*\c-\c) rectangle ++(\c,\c);
    \node[dlab] at (0.65+\j*\c+0.5*\c,-0.58-\i*\c-0.5*\c) {\s};
  }
}
\node[dlab, align=center] at (0.97,-2.3) {Equal numbers,\\equal weight};
\node[dlab, anchor=south] at (1.31,-0.33) {Matrix limb};
\node[dlab, rotate=90, anchor=south] at (0.36,-1.24) {Key limb};
\foreach \t [count=\k] in {{Add equal-weight limb sums into 7 sums},
                           {Weight each sum into one 64-bit total},
                           {Fold the upper 32 bits back in, twice},
                           {Subtract the prime once if needed}} {
  \node[dbox, fill=apPetrolL!60, minimum width=5.2cm, minimum height=0.3cm] (r\k) at (5.65,-0.4-\k*0.56) {\t};
}
\foreach \k [evaluate=\k as \n using int(\k+1)] in {1,2,3} \draw[dgo] (r\k.south) -- (r\n.north);
\draw[dgo] (2.0,-0.96) -- (r1.west);
\end{tikzpicture}%
}
\caption{Column-group packing. (a) Each output column depends on one matrix column. Each core computes and recombines whole groups of 32 columns. (b) The 16 limb sums of an output share seven weights.}
\label{fig:pack}
\vspace{-3mm}
\end{figure}

\noindent\textbf{Reading the packing matrix once.}
Packing follows the same principle as the scan.
The packing matrix is large and shared by all clients, whereas a packing key holds only 8,192 values regardless of the number of records.
For example, the packing matrix takes 1\,GB for 32\,KB records, while the keys of 128 clients take only 4\,MB.
Therefore, \sys reads the packing matrix from memory once per batch and keeps the keys of the batch in the shared L3 cache.
Every core reads all keys, and those of a large batch exceed the L2 cache of a core.

\noindent\textbf{One column group at a time.}
Figure~\ref{fig:pack}~(a) follows core~1.
\textcircled{\footnotesize 1}~It loads the slices of its group from the four byte planes of the packing matrix into its L2 cache.
\textcircled{\footnotesize 2}~The accelerator multiplies them by the packing keys of the batch from the L3 cache and adds the products into the 16 limb sums of each client and column, kept in the L2 cache.
When the group is done, \textcircled{\footnotesize 3}~the core recombines them into the 32 outputs of every client and \textcircled{\footnotesize 4}~moves to the next group.
Within a core, the matrix slice stays in the L2 cache while the keys of all clients pass through.

\noindent\textbf{Exact recombination without division.}
Recombining right away is cheap because of two arithmetic properties, shown in Figure~\ref{fig:pack}~(b).
First, the weight of a limb sum depends only on the sum of the positions of its two limbs.
Each cell of the grid in the figure shows this sum for one key limb and one matrix limb.
For example, key limb~1 times matrix limb~2 has the same weight $2^{24}$ as key limb~2 times matrix limb~1.
Thus, the 16 limb sums of an output fall into seven groups of equal weight.
The engine adds up each group and multiplies only seven sums by their weights.

Second, the prime of the scheme is $2^{32}-12{,}287$, just below $2^{32}$.
Therefore, $2^{32}$ equals 12,287 modulo the prime, and the part of a sum above 32 bits folds into only 12,287 times its value.
For example, $3\cdot2^{32}+7$ folds into $3\cdot12{,}287+7=36{,}868$.
This keeps the weighted total of an output within one 64-bit lane.
Two folds of its upper half and one conditional subtraction of the prime give the exact result, as the four steps of the figure show.
All of these steps are vector operations on registers. Consequently, the recombination needs neither a division nor a pass through memory.

\subsection{Splitting Large Batches \label{design:placement}}

A larger batch allows each database byte to serve more queries, and the scan should keep its speed as the batch grows.
However, the query limbs of a block must fit in half of the L2 cache (\S\ref{design:scan}).
Therefore, a block holds fewer records when the batch holds more clients.
Doubling the batch halves the block length.
The scan then runs twice as many blocks, and each block recombines twice as many clients.
As a result, the recombinations grow fourfold.
\sys instead splits a large batch into \emph{sub-batches} of 128 clients and runs the scan once per sub-batch.
The block length then stays fixed, and the recombinations grow only in proportion to the batch size.

\begin{figure}[t]
\centering
\begin{tikzpicture}[x=1cm,y=1cm]
\def\x{2.15} \def\b{0.09375}
\draw[dtile, fill=apStoneL] (\x,1.25) rectangle ({\x+6},1.50);
\fill[apStone!55] (\x,1.25) rectangle ({\x+\b},1.50);
\foreach \k in {1,...,63} \draw[apStone!80, line width=0.2pt] ({\x+\k*\b},1.25) -- ++(0,0.25);
\draw[dtile] (\x,1.25) rectangle ({\x+6},1.50);
\draw[dtile, fill=apLimb!30] (\x,1.50) rectangle ({\x+\b},2.70);
\draw[apPetrol, line width=0.9pt] (\x,1.25) rectangle ({\x+\b},2.70);
\node[dlab, anchor=east, align=right] at ({\x-0.12},1.98) {One batch\\of 512 clients};
\node[dlab, anchor=south west, align=left] at ({\x+0.22},1.56) {Half of L2 holds 512 clients $\times$ 512 records\\64 blocks, each recombining 512 clients};
\draw[dtile, fill=apStoneL] (\x,0.10) rectangle ({\x+6},0.35);
\fill[apStone!55] (\x,0.10) rectangle ({\x+4*\b},0.35);
\foreach \k in {1,...,15} \draw[apStone!80, line width=0.35pt] ({\x+\k*4*\b},0.10) -- ++(0,0.25);
\draw[dtile] (\x,0.10) rectangle ({\x+6},0.35);
\draw[dtile, fill=apLimb!30] (\x,0.35) rectangle ({\x+4*\b},0.65);
\draw[apPetrol, line width=0.9pt] (\x,0.10) rectangle ({\x+4*\b},0.65);
\node[dlab, anchor=east, align=right] at ({\x-0.12},0.42) {Four sub-batches\\of 128 clients};
\node[dlab, anchor=south west, align=left] at ({\x+4*\b+0.22},0.40) {Half of L2 holds 128 clients $\times$ 2,048 records\\16 blocks, each recombining 128 clients};
\end{tikzpicture}%
\caption{Sub-batches, drawn to scale. Each bar is a database of 32,768 records, and each box is the query limbs of one block, which fill half of the L2 cache. Four sub-batches of 128 clients keep the blocks four times as long as one batch of 512 does.}
\label{fig:placement}
\end{figure}
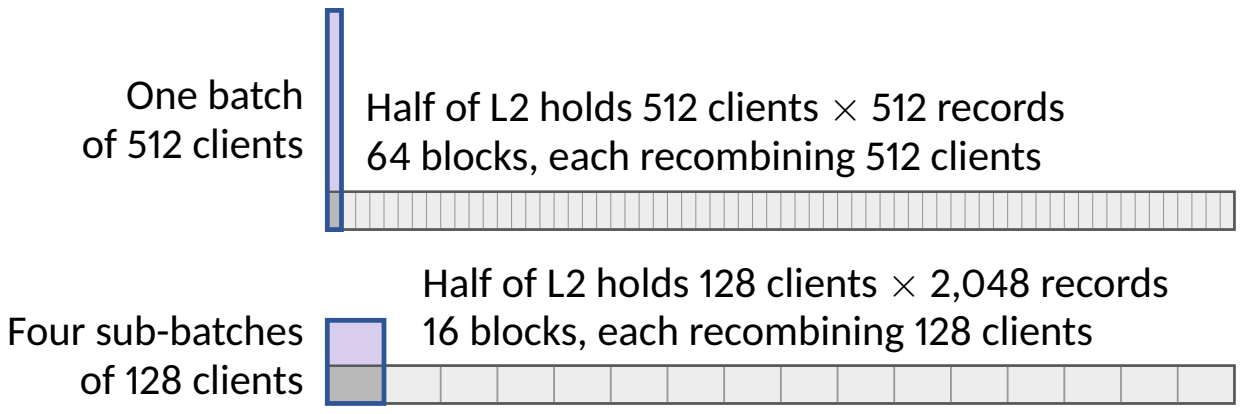

Figure~\ref{fig:placement} shows 512 clients over 32,768 records.
As one batch, they would run 64 blocks that each recombine 512 clients.
As four sub-batches, they run 16 blocks per sub-batch that each recombine 128 clients, a quarter of the recombinations.
Each sub-batch reads the database once more.
This costs little, because a sub-batch of 128 clients already makes the scan bound by the accelerator rather than by memory.
Sub-batches matter only when the maximum batch size exceeds 128.
Packing is not split, because it has no block that shrinks with the batch.

\subsection{In-Place Engine and Vectorized Tail \label{design:engine}}

Request handling should not leave the accelerator idle, but it takes longer than the scan and packing would at the peak rate (\S\ref{sec:motiv-gap}).
The cores handle each request at two points.
Before the scan and packing, they check the request and convert its query vector and two keys into the form that these products read.
After the products, they run the tail, which turns the outputs of the scan and packing into the response of each client.
A response holds several polynomials, and the intermediate results of the tail take tens of megabytes per batch.
We observe that a request already stores the query vector and keys as little-endian 32-bit integers, the form in which the scan and packing read them.
Therefore, a request needs only to be checked, never decoded or copied.
Similarly, the tail of one client and one response polynomial needs only its own parts of the scan and packing outputs and fits in the L1 cache.
As a result, \sys validates requests in place and runs one vectorized tail per such pair.

\begin{figure}[t]
\centering
\subfloat[Frames read in place]{\begin{tikzpicture}[x=1cm,y=1cm,
  pnum/.style={circle, draw=apEdge, line width=0.45pt, fill=apPetrolL, inner sep=0pt, minimum size=3.3mm, font=\figfontS}]
\fill[apStone!22] (0,2.15) rectangle (0.95,2.75);
\fill[apStone!22] (0,0.05) rectangle (0.95,2.1);
\node[dbandlab] at (0.475,2.45) {Core};
\node[dbandlab] at (0.475,1.075) {Memory};
\fill[apCore] (1.0,2.15) rectangle (8.4,2.75);
\fill[apStoneL] (1.0,0.05) rectangle (8.4,2.1);
\foreach \y in {1.45,0.65} {
  \draw[dtile, fill=white] (1.1,\y) rectangle (1.45,\y+0.34);
  \draw[dtile, fill=apLimbL] (1.45,\y) rectangle (3.85,\y+0.34);
  \draw[dtile, fill=apPetrolL] (3.85,\y) rectangle (5.45,\y+0.34);
  \draw[dtile, fill=apCacheD] (5.45,\y) rectangle (6.65,\y+0.34);
}
\node[dlab] at (1.275,1.62) {H};
\node[dlab] at (2.65,1.62) {Query vector};
\node[dlab] at (4.65,1.62) {Packing key};
\node[dlab] at (6.05,1.62) {Tail key};
\draw[dtile, fill=apBrickL] (1.1,1.05) rectangle (6.65,1.39);
\node[dlab] at (3.875,1.22) {Malformed frame, removed from the batch};
\node[dlab] at (3.875,0.33) {Request frames};
\foreach \y in {1.45,1.05,0.65} \draw[dtile, fill=white] (6.85,\y) rectangle (8.3,\y+0.34);
\node[dlab] at (7.575,0.33) {Responses};
\node[dbox, fill=apPetrolL, minimum height=0.4cm] (chk) at (1.6,2.45) {Check};
\node[dbox, fill=apPetrolL, minimum height=0.4cm] (prd) at (3.65,2.45) {Scan, packing};
\node[dbox, fill=apPetrolL, minimum height=0.4cm] (tl) at (6.6,2.45) {Vector tail};
\draw[dgo] (1.275,1.79) -- (1.275,2.25);
\draw[dgo] (2.9,1.79) -- (2.9,2.25);
\draw[dgo] (4.4,1.79) -- (4.4,2.25);
\draw[dgo] (6.05,1.79) -- (6.05,2.25);
\draw[dgo] (prd) -- (tl);
\draw[dgo] (7.35,2.25) -- (7.35,1.79);
\node[pnum] at (1.53,2.02) {1};
\node[pnum] at (3.65,2.02) {2};
\node[pnum] at (6.3,2.02) {3};
\node[pnum] at (7.6,2.02) {4};
\end{tikzpicture}%
}\\
\subfloat[Tail of one client and response polynomial]{\begin{tikzpicture}[x=1cm,y=1cm]
\fill[apStone!22] (0,0.75) rectangle (0.95,1.65);
\fill[apStone!22] (0,0.05) rectangle (0.95,0.7);
\node[dbandlab] at (0.475,1.2) {L1 cache};
\node[dbandlab] at (0.475,0.375) {Memory};
\fill[apCache] (1.0,0.75) rectangle (8.4,1.65);
\fill[apStoneL] (1.0,0.05) rectangle (8.4,0.7);
\foreach \t [count=\k] in {Add key\\term, Inverse\\NTT, Add scan\\output, Switch\\modulus, Write\\bits}
  \node[dbox, fill=apPetrolL, minimum width=1.05cm, minimum height=0.6cm] (z\k) at (0.45+\k*1.45,1.2) {\t};
\foreach \k [evaluate=\k as \n using int(\k+1)] in {1,...,4} \draw[dgo] (z\k) -- (z\n);
\draw[dtile, fill=apPetrolL] (1.1,0.16) rectangle (2.75,0.52);
\node[dlab] at (1.9,0.34) {Packing part};
\draw[dtile, fill=apLimbL] (4.05,0.16) rectangle (5.55,0.52);
\node[dlab] at (4.8,0.34) {Scan part};
\draw[dtile, fill=white] (6.9,0.16) rectangle (8.3,0.52);
\node[dlab] at (7.625,0.34) {Response};
\draw[dgo] (1.9,0.52) -- (z1.south -| 1.9,0);
\draw[dgo] (4.8,0.52) -- (z3.south -| 4.8,0);
\draw[dgo] (z5.south -| 7.6,0) -- (7.6,0.52);
\end{tikzpicture}%
}
\caption{Request handling with few core cycles. (a) The cores check each request where it arrived, and the scan and packing read it there without a copy. (b) The tail of one pair runs in vector registers and the L1 cache.}
\label{fig:engine}
\end{figure}
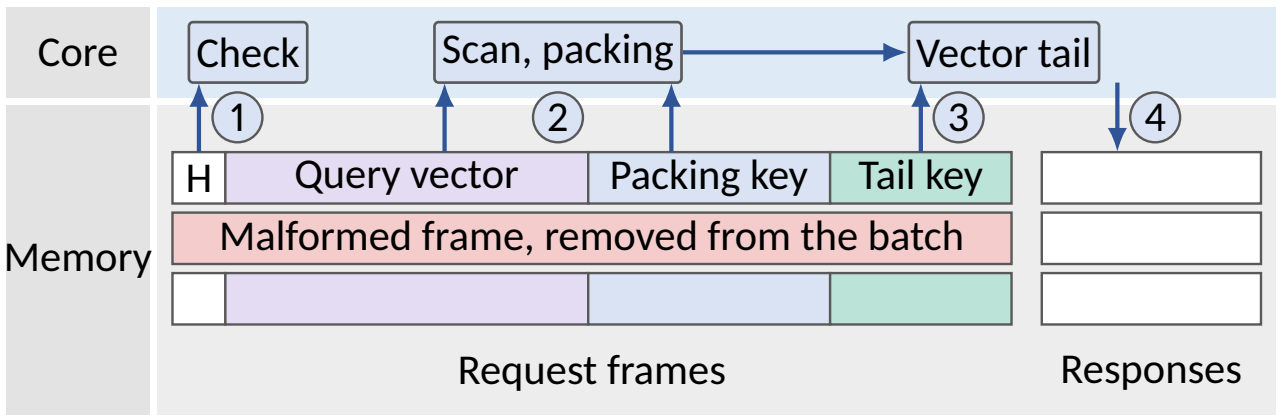
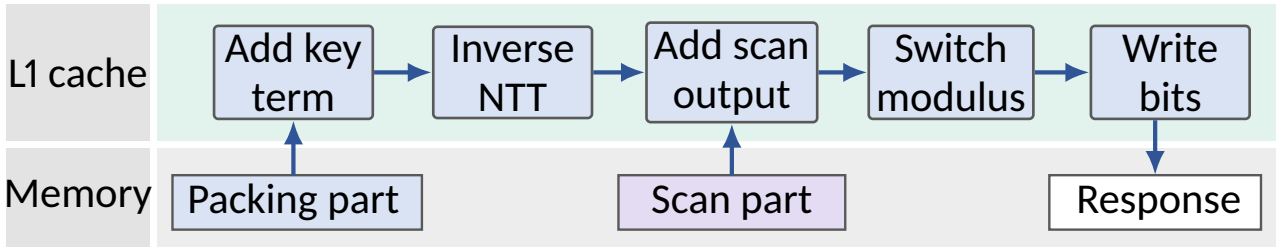

\noindent\textbf{Validating requests in place.}
\sys answers requests in the frames where they arrive, as Figure~\ref{fig:engine}~(a) shows.
\textcircled{\footnotesize 1}~The engine checks the lengths, the headers, and the range of every key value of the received frames.
It runs these checks with vector instructions, in parallel over the whole batch.
A frame that fails a check is removed from the batch and receives an error status.
For example, if one frame of a batch of 128 carries a key value at or above the prime, the engine answers the other 127.
The check accepts and rejects exactly the frames that a conventional decoder does.
Then \textcircled{\footnotesize 2}~the scan and packing and \textcircled{\footnotesize 3}~the tail read the query vectors and keys directly from the frames.
Their intermediate results are kept in buffers that the engine allocates once and reuses for every batch.

\noindent\textbf{One vectorized tail per pair.}
The tail runs as one parallel loop over all pairs of a client and a response polynomial.
For example, a batch of 128 clients with 32\,KB records gives 2,048 pairs, because each response holds 16 polynomials.
Packing yields its output in the number-theoretic transform (NTT) form, in which the scheme multiplies polynomials.
For each pair, the tail adds a term computed from the second key of the client and converts the sum back with the inverse NTT.
It then adds the scan output, switches the modulus, and \textcircled{\footnotesize 4}~writes the result as the second row of the response ciphertext.
The first row is the same for all clients and is precomputed offline (\S\ref{design:overview}).
All intermediate values of one pair fit in tens of kilobytes and stay in the L1 cache, as Figure~\ref{fig:engine}~(b) shows.
Every step processes eight coefficients at a time with AVX-512 instructions.
Every remainder modulo the prime comes from the two folds of packing instead of a division.

\subsection{Discussion \label{design:discussion}}

\noindent\textbf{Security.}
\sys keeps the threat model of \sandwich~\cite{sandwichpir}, whose scheme it follows.
The server is honest but curious about which records the clients fetch.
\sys does not protect the database from the clients.
Because \sys computes exactly the responses of this scheme, its security analysis still holds.
Batching reveals nothing new, because the server already sees which requests arrive together, and the batch time depends only on public sizes (i.e., the number of requests and the database shape).
A malformed request cannot change another response, because each response goes only into the buffer of its own request.
Removing a malformed request changes only the number of requests.

\noindent\textbf{Database updates.}
When a record changes, the offline state that \sys derives from the database changes too.
This state comes from the hint, which is linear in the database. Therefore, changing one record adds one outer product to the hint.
The rest of the state depends on the hint but not on the number of records.
\sys recomputes it from the updated hint while it rewrites only the changed records.
Meanwhile, \sys answers batches with the old state.
Clients need no action, because they keep no state.
However, recomputing the state costs about the same for one changed record as for many. Therefore, updates should be applied in groups.

\noindent\textbf{Key-value lookups.}
Many applications look up records by key, while the scheme fetches them by index.
A lookup by key reduces to fetching by index~\cite{chorkeyword}.
Each pair is stored in one of two records chosen by public hash functions, as in cuckoo hashing~\cite{cuckoohashing}.
A client fetches both records and keeps the one that holds its key.
Therefore, every lookup sends two queries whether or not the key is present, and the server learns nothing new.
However, lookups then run at half the query rate, and the hashing leaves some records empty.
Records that each hold several pairs under one hash function can bring a lookup back to one query, at the cost of larger records.
Inserting a pair may move a few other pairs, each a record update.

\noindent\textbf{Other on-CPU accelerators.}
IBM Power10 and recent Arm CPUs also include matrix accelerators (\S\ref{sec:motiv-why}).
Like AMX, both accumulate 8-bit products into 32-bit sums~\cite{power10mma,armsme}.
Therefore, the limb split and the recombination in the scan and packing carry over unchanged.
Both also read their operands through the caches of the core, and our design principle and sub-batches therefore apply as well.

However, a port must change three parts.
First, both accelerators take their operands from vector registers instead of tiles.
The scan would load the query limbs into vector registers and set its blocking by the accumulators, which are eight blocks of $4\times4$ sums on Power10~\cite{power10mma}.
Second, the 8-bit products of Power10 take one signed and one unsigned operand.
Packing already stores its matrix as signed bytes, and the scan would shift one operand by 128 and add back an exact correction for each output.
Third, in the SME CPUs, one SME unit serves a cluster of cores~\cite{smepilot}.
The scan would then run one thread per unit.
The tail would also move from AVX-512 to the vector instructions of each CPU.
We leave these ports to future work.

\noindent\textbf{Other matrix-shaped schemes.}
Other hintless schemes~\cite{hintlesspir,ypir,inspire,reinspire} also reduce most of their server computation to integer products with an operand that all clients share.
However, \inspire and \reins use moduli~\cite{reinspire} wider than the 32-bit values of \sandwich.
When the modulus is a power of two, as in \reins, the query-blocked scan applies to such products as well, with more byte limbs per value and wider accumulators per block.
When the modulus is made of NTT-friendly primes, as in \inspire~\cite{reinspire}, a scheme could instead recombine each output inside the core, as column-group packing does for its prime.
In both cases, the extra limb products cost more at small batches.
Furthermore, a whole scheme gains only if its per-client steps are batched too, as \sys does for packing.

\section{Implementation \label{implementation}}
We build \sys on the codebase of \sandwich~\cite{sandwichpir}.
From it, \sys takes the parameters, key generation, and message formats of the scheme.
The engine of \sys is new and consists of about 600 lines of C++ with AMX and AVX-512 intrinsics and OpenMP.
The server, client, and benchmark programs of \sys add about 1,500 lines of Rust.
The scan uses \texttt{TDPBUUD} for unsigned limb products.
Packing uses \texttt{TDPBUSD} with the packing matrix stored as signed bytes, which \texttt{VPDPBUSD} of AVX-512 VNNI can read in the same layout.
\sys keeps freed memory in the C allocator with \texttt{mallopt}, so that the next batch takes no new page faults on it.
It uses a length-prefixed binary protocol over TCP.
The client uses the original library and is compiled for AVX2, because our client machines lack AVX-512.
All programs are built with rustc~1.95.0 and GCC~13.3.

\section{Evaluation \label{evaluation}}

\subsection{Methodology \label{sec:eval-method}}

\noindent\textbf{Testbed.}
We use a server with two Intel Xeon Silver 4514Y CPUs and 256\,GB of DDR5 memory.
Each CPU has 16 cores, 2\,MB of L2 cache per core, and 30\,MB of L3 cache.
Four client machines with Intel Core i5-12600K CPUs connect to the server through a 100\,GbE switch.
All machines run Ubuntu~24.04 with Linux~6.17.
Unless specified, every system uses only one CPU to avoid NUMA effects.

\noindent\textbf{Compared schemes.}
We compare \sys with the state-of-the-art hintless schemes \reins~\cite{reinspire}, \inspire~\cite{inspire}, \valar~\cite{valar}, and \ypirsp~\cite{ypir}, and with the hint-based \simplepir~\cite{simplepir} as a reference.
Each runs with its default parameters.
These give about 128 bits of security by the authors' estimates, except for \valar, whose authors make no such claim.
We run each in its fastest configuration, and \ypirsp runs only on one core.
\reins uses the 54-bit modulus of its paper and a 64-bit index that its implementation needs at 32\,GB.
In addition, we compare \port, our optimized AMX system from \S\ref{sec:motiv-gap}.
As a GPU-based system, we run \sandwich~\cite{sandwichpir} on one NVIDIA A30 GPU with 24\,GB of memory in a CloudLab server~\cite{cloudlab}.
It runs consecutive batches at its best batch size, which allows the GPU to run at its highest clock.

\begin{figure}[t]
\centering
\includegraphics[width=0.68\linewidth]{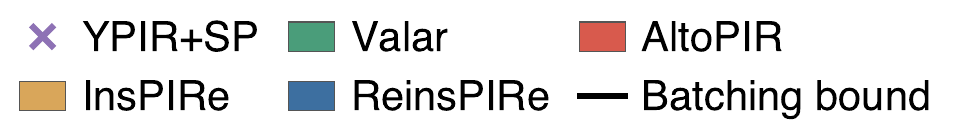}\\
\subfloat[Queries per second]{\includegraphics[width=0.48\linewidth]{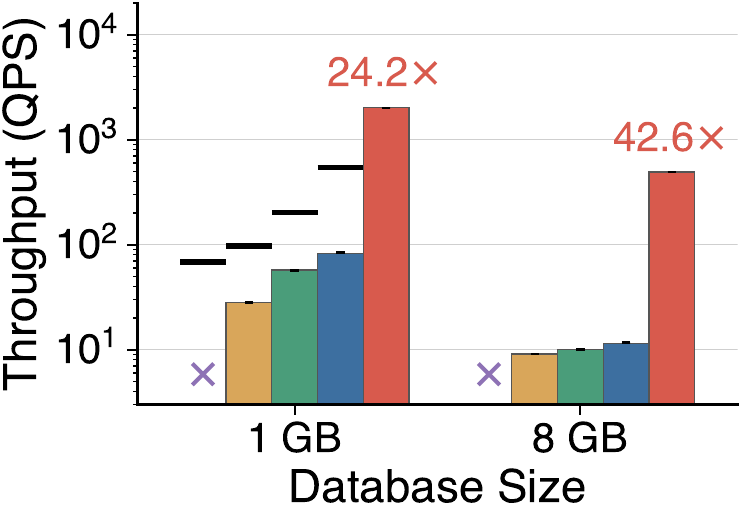}}\hfill
\subfloat[Database bytes per second]{\includegraphics[width=0.48\linewidth]{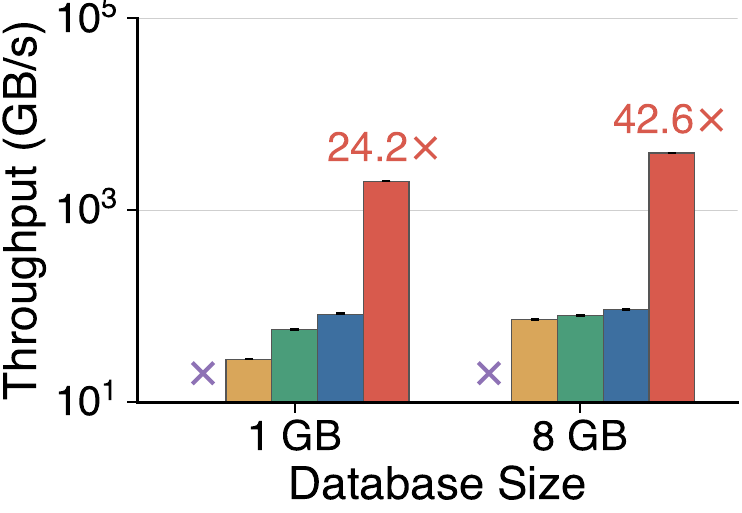}}
\caption{\sys answers far more queries than existing hintless schemes. Black marks show each existing scheme at 1\,GB if its pass over the database took no time. \ypirsp runs only on one core.}
\label{fig:compare}
\end{figure}

\noindent\textbf{Workloads and parameters.}
The databases hold 32\,KB records and range from 1\,GB ($2^{15}$ records) to 64\,GB ($2^{21}$ records).
Every query comes from a new client with fresh keys and a random index, and every response is decrypted and compared with the record.
\sys follows the parameters of \sandwich~\cite{sandwichpir}, for which \sandwich reports at least 128 bits of security.
On 64\,GB, each existing scheme uses its fastest configuration on 32\,GB.

\noindent\textbf{Measurement.}
For throughput, we give each system a batch of requests that have already arrived and divide the number of requests by the time from the received request bytes to the serialized response bytes.
We run each configuration in at least three processes after a warmup and report the median of the per-process medians.
In the arrival experiments, the four clients send requests as a Poisson process for 20 seconds over eight connections.
A request that arrives when 512 requests are waiting receives a busy reply.
The latency runs from the scheduled send time to the receipt of the full response.
In these experiments, the engine of \sys runs on 15 cores, and all reader and writer threads share the single remaining core.

\subsection{Main Results \label{sec:eval-main}}

\begin{figure}[t]
\centering
\includegraphics[width=0.75\linewidth]{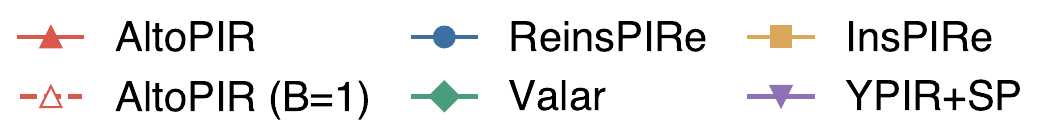}\\
\subfloat[16 cores]{\includegraphics[width=0.48\linewidth]{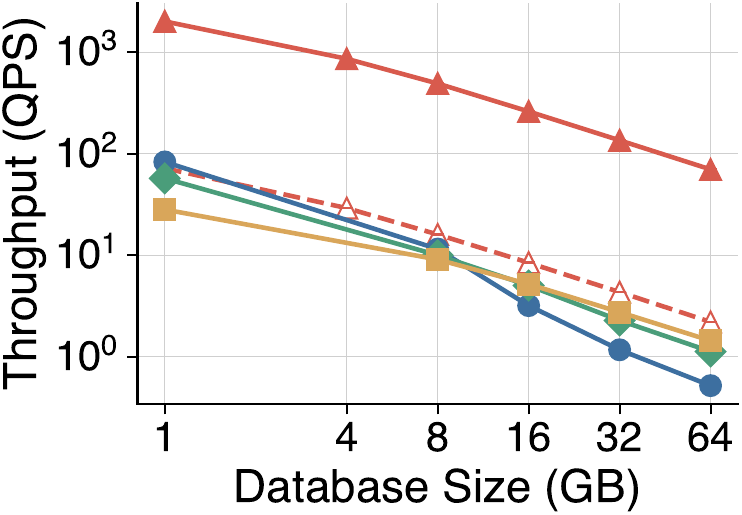}}\hfill
\subfloat[1 core]{\includegraphics[width=0.48\linewidth]{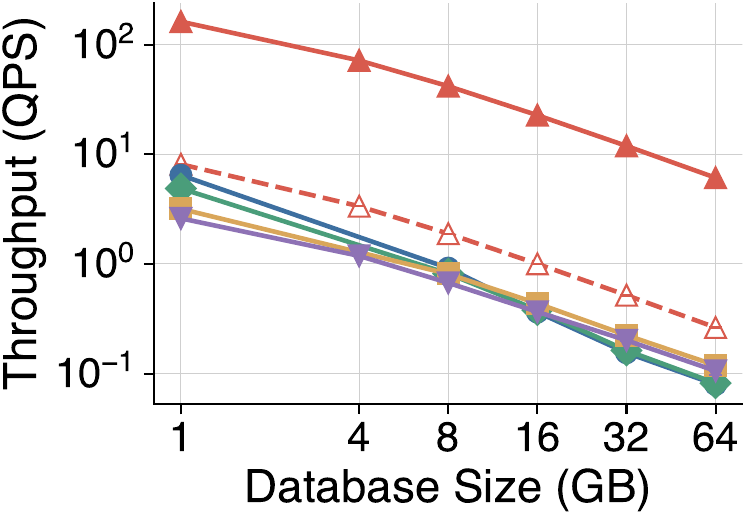}}
\caption{\sys keeps its advantage at every database size. \ypirsp runs only on one core.}
\label{fig:dbsize}
\end{figure}

\noindent\textbf{Throughput on 1\,GB and 8\,GB databases.}
Figure~\ref{fig:compare} shows that \sys outperforms every existing scheme.
It outperforms the fastest of them, \reins, by 24.2$\times$ on 1\,GB and by 42.6$\times$ on 8\,GB.
This is because \sys batches both the scan and packing on the accelerator, while the existing schemes batch at most the scan and run packing for each client.
Figure~\ref{fig:compare}~(b) counts one pass over the database per answered query and shows that the throughput of \sys in database bytes per second exceeds the memory bandwidth by more than an order of magnitude.
The black marks assume that the pass of each existing scheme over the database takes no time.
Even so, \sys would still outperform the highest of these marks by 3.71$\times$, because of the per-client steps of these schemes.
This shows that packing must be batched as well as the scan.

\noindent\textbf{Impact of database size.}
Figure~\ref{fig:dbsize}~(a) shows that the advantage of \sys over the fastest existing scheme grows with the database size, from 24.2$\times$ at 1\,GB to 50.1$\times$ at 16\,GB.
The advantage grows because packing and the tail take about half of a batch of \sys on 1\,GB, and neither grows with the database.
In contrast, \reins spends most of each query in its pass over the database.
At 16\,GB and above, the scan takes more than 90\% of a batch of \sys as well, and the advantage stops growing.
On one core, \sys with one query per batch also outperforms the fastest existing scheme by 1.26$\times$ to 2.32$\times$, as Figure~\ref{fig:dbsize}~(b) shows.
Overall, the larger the database, the more servers \sys saves.

\begin{table}[t]
\centering
\small
\caption{Comparison on a 1\,GB database.\label{tab:compare}}
\setlength{\tabcolsep}{1.8pt}
\begin{threeparttable}
\begin{tabular}{l r r r r r r r}
\toprule
 & \multicolumn{2}{c}{Comm. (KB)} & Client & \multicolumn{2}{c}{Latency (ms)} & Tput & Offline \\
\cmidrule(lr){2-3}\cmidrule(lr){5-6}
Scheme & Up & Down & (ms) & 1 core & 16 cores & (QPS) & (s) \\
\midrule
\simplepir & 64 & 211\tnote{a} & 57.3 & 76.5 & -- & -- & 499 \\
\ypirsp & 686 & 120 & 52 & 384 & -- & -- & 14.9 \\
\inspire & 392 & 96 & 40.3 & 315 & 315 & 28.3 & 28.5 \\
\valar & 280 & 152 & 33.1 & 205 & 20.6 & 57.3 & 3.0 \\
\reins & 270 & 94 & 63.5 & 155 & 155 & 83.2 & 101 \\
\sys & 192 & 112 & 18.0 & 123 & 13.9 & 2{,}013 & 13.4 \\
\bottomrule
\end{tabular}
\begin{tablenotes}\footnotesize
\item[a] Each client also downloads a 211\,MB hint in advance.
\end{tablenotes}
\end{threeparttable}
\end{table}

\noindent\textbf{Latency, communication, and client cost.}
Table~\ref{tab:compare} compares the cost of one query on a 1\,GB database.
\sys achieves the lowest latency among the hintless schemes on both one core and 16 cores.
The client of \sys needs the least time and uploads the fewest bytes among the hintless schemes.
In short, \sys raises the throughput without making a single query slower or more costly for its client.

\begin{figure}[t]
\centering
\includegraphics[width=\linewidth]{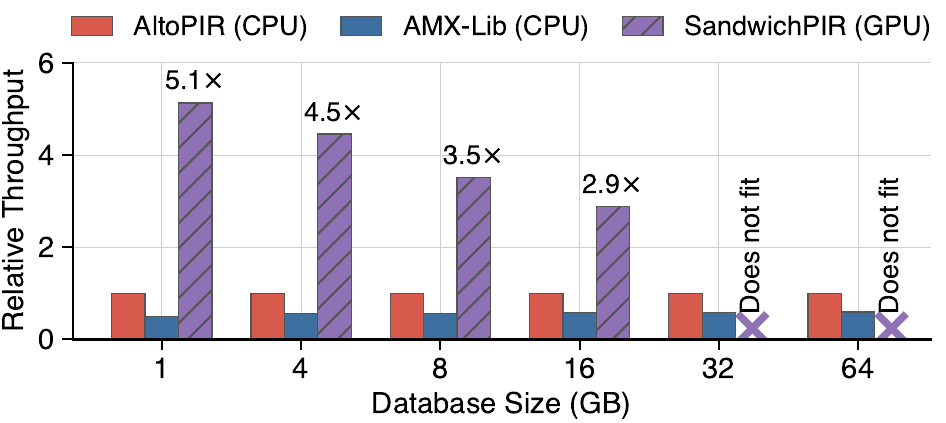}
\caption{\sys and \port against \sandwich on one NVIDIA A30 GPU, in throughput relative to \sys. The GPU answers more queries per second, but the two largest databases do not fit in its memory.}
\label{fig:gpu}
\end{figure}

\noindent\textbf{Comparison with GPU-based systems.}
Figure~\ref{fig:gpu} shows that \sandwich on the GPU achieves higher throughput than \sys by 2.9$\times$ to 5.1$\times$.
This is expected from the rated multiply-add rate and memory bandwidth of the GPU~\cite{a30datasheet}.
Databases of 32\,GB and 64\,GB do not fit in the 24\,GB memory of the GPU, while \sys serves them.
This limit is not specific to the A30.
Even the largest single GPUs hold a few hundred gigabytes, while the main memory of a CPU server reaches several terabytes.
\sys also outperforms \port by 1.68$\times$ to 2.00$\times$ at every size.
Therefore, \sys complements GPU-based systems on CPU servers and on databases beyond the memory of one GPU.

\begin{figure}[t]
\centering
\subfloat[Latency of \sys]{\includegraphics[width=0.48\linewidth]{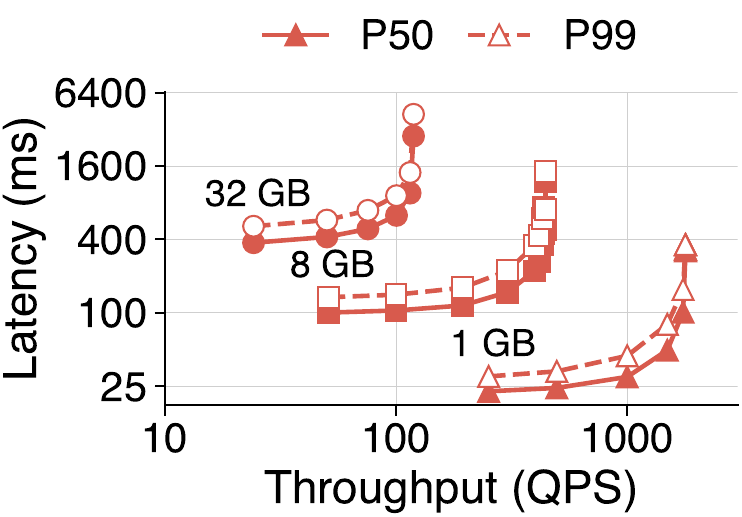}}\hfill
\subfloat[Maximum batch size and VNNI (1\,GB)]{\includegraphics[width=0.48\linewidth]{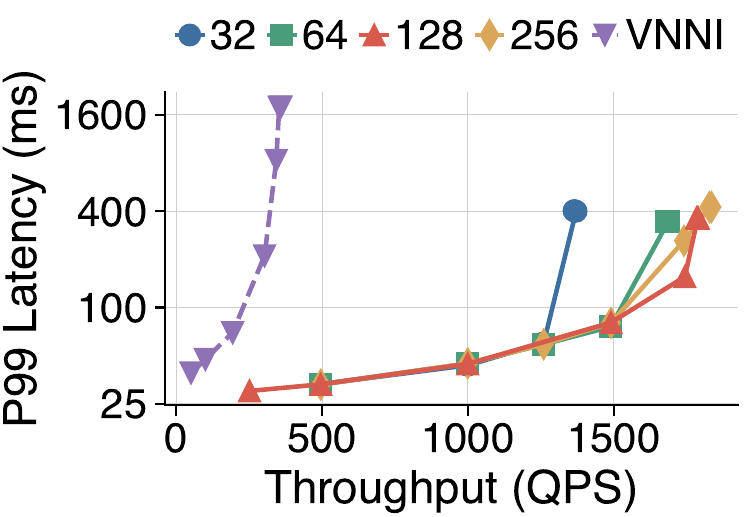}}
\caption{Latency of \sys under Poisson arrivals. The latency stays low until the load nears saturation.}
\label{fig:openloop}
\end{figure}

\noindent\textbf{Latency under load.}
Figure~\ref{fig:openloop}~(a) shows the latency of \sys when requests arrive at Poisson rates.
On 1\,GB, \sys answers 99\% of the requests within about 30\,ms at 250\,QPS, three times the 83\,QPS that \reins sustains at most.
To reach those 83\,QPS, \reins answers batches of 64 queries in 769\,ms each, and every query of a batch waits for the whole batch.
\vnni runs \sys with AVX-512 VNNI instructions instead of AMX.
At about 340\,QPS, close to the most that \vnni sustains, its 99th-percentile latency reaches 831\,ms, as Figure~\ref{fig:openloop}~(b) shows.
\sys keeps it at 30\,ms at 250\,QPS and 33\,ms at 500\,QPS, below and above that load.
Thus, the accelerator keeps the latency low at loads where the other systems have already saturated.

\subsection{End-to-End Application Performance \label{sec:eval-apps}}

We now demonstrate that \sys can speed up leaked-password checks, unsafe-link checks, and privacy-preserving RAG.
We report latency when 64 lookups arrive at once.
It includes one round trip of 23\,ms and the transfer time on the mobile network that \sandwich assumes, 112.07\,Mbps down and 14.68\,Mbps up like prior work~\cite{sandwichpir}.

\begin{figure*}[t]
\centering
\includegraphics[scale=0.375]{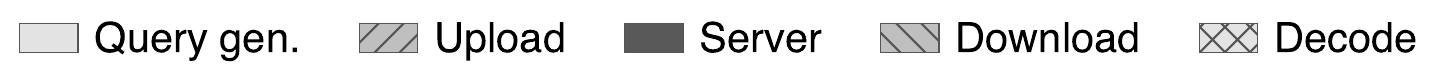}\\
\subfloat[Passwords: lookups per second]{\includegraphics[width=0.24\linewidth]{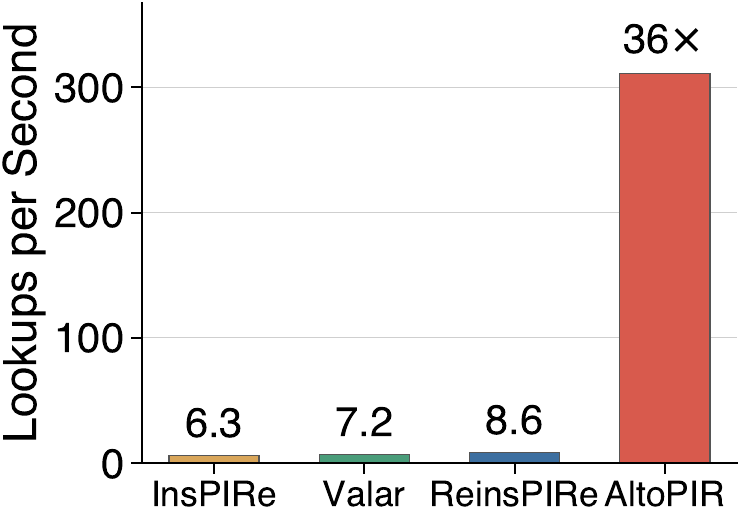}}\hfill
\subfloat[Passwords: 64 lookups at once]{\includegraphics[width=0.24\linewidth]{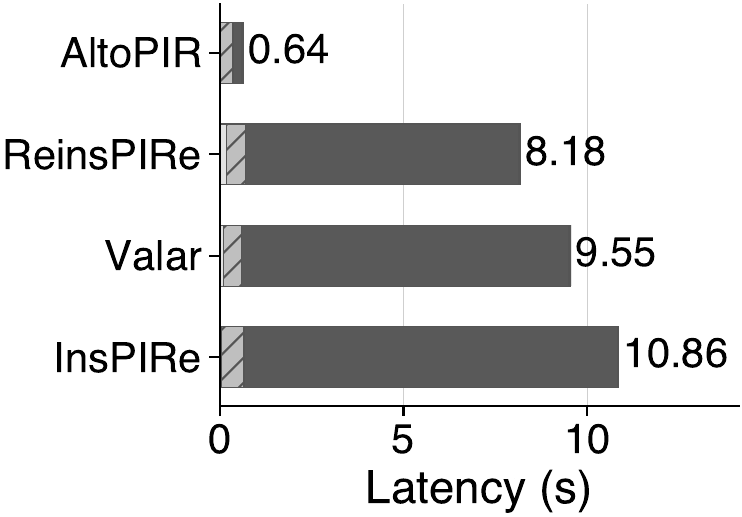}}\hfill
\subfloat[Links: lookups per second]{\includegraphics[width=0.24\linewidth]{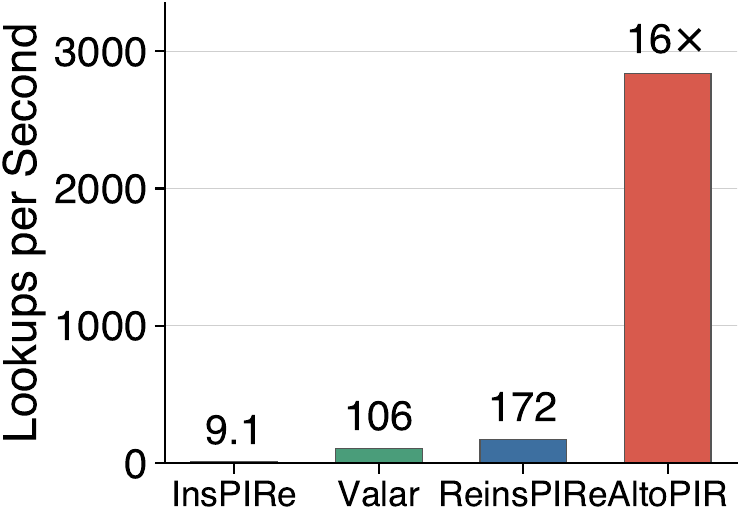}}\hfill
\subfloat[Links: 64 lookups at once]{\includegraphics[width=0.24\linewidth]{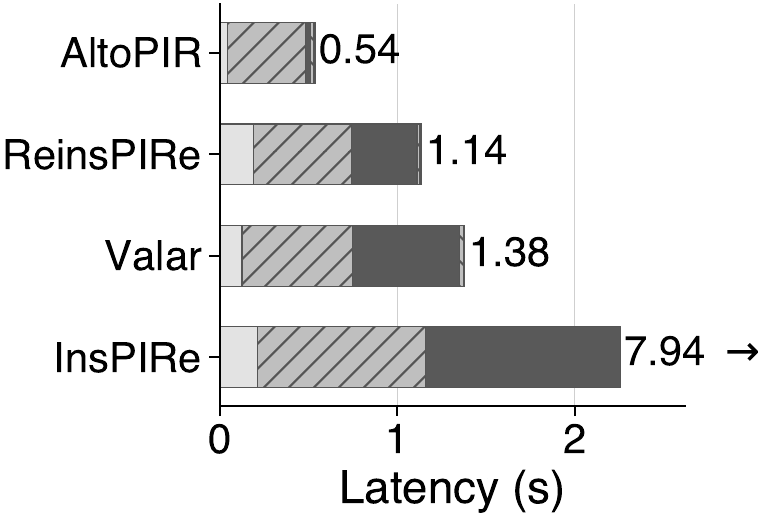}}
\caption{Leaked-password and unsafe-link checks. \sys answers more lookups per second and keeps a lookup fast when 64 lookups arrive at once over a mobile network.}
\label{fig:apps}
\end{figure*}

\noindent\textbf{Leaked-password checks.}
Checking a password against breach databases is one of the most widely deployed private lookups.
Deployed services receive a prefix of the password hash, which still helps an attacker guess the password~\cite{c3checks}.
We place the SHA-1 hashes of the 2.07 billion leaked passwords of Have I Been Pwned~\cite{hibp} in $2^{17}$ buckets of 88\,KB (11\,GB).
A check fetches the bucket of its hash.
Figure~\ref{fig:apps}~(a) shows that \sys achieves higher throughput than \reins, the fastest existing scheme, by 36.1$\times$.
When 64 checks arrive at once, \sys achieves better latency than the existing schemes by 12.8$\times$ to 17.0$\times$, as Figure~\ref{fig:apps}~(b) shows.
This is because \sys answers all 64 requests together, faster than \reins by 33.9$\times$.
As a result, many users can check their passwords at the same time without a long wait.

\noindent\textbf{Unsafe-link checks.}
Browsers screen the links that users open against blocklists of unsafe sites, and Safe Browsing reveals a hash prefix of each link that matches its local list~\cite{checklist}.
The database holds the 1.1 million phishing links and domains of Phishing.Database~\cite{phishingdb} in $2^{14}$ buckets of 4\,KB (64\,MB).
A check sends six requests for the link, its host, and its parent domains.
Random requests pad every link to six.
Figure~\ref{fig:apps}~(c) shows that \sys provides better throughput than \reins by 16.4$\times$.
When 64 links arrive at once, \sys answers all 384 requests faster than \reins by 14.6$\times$, as Figure~\ref{fig:apps}~(d) shows.
Therefore, \sys also helps small databases that need several requests per lookup.

\begin{figure}[t]
\centering
\includegraphics[width=\linewidth]{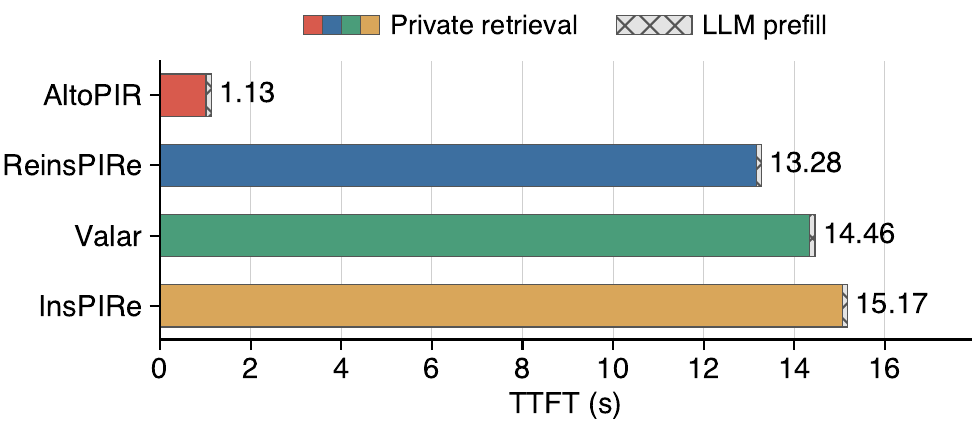}
\caption{Time to first token (TTFT) of a RAG question when 64 questions arrive at once. \sys reduces the wait to about one second.}
\label{fig:rag}
\end{figure}

\noindent\textbf{Privacy-preserving RAG.}
Retrieval-augmented generation grounds the answer of a language model in passages fetched for each question~\cite{rag}.
Recent systems find these passages without revealing the question~\cite{pirrag,pacmann,compass}.
We group the 21 million Wikipedia passages of DPR~\cite{dpr} by their BGE embeddings~\cite{cpack} into $2^{18}$ clusters of 64\,KB (16\,GB).
For each test question of Natural Questions~\cite{nq}, the client fetches the closest cluster.
Llama~3.1~8B~\cite{llama3} then reads its five closest passages on one NVIDIA A30 GPU.
Figure~\ref{fig:rag} shows that \sys achieves a shorter time to first token (TTFT) than the existing schemes by 11.8$\times$ to 13.4$\times$.
This is because their private retrieval takes 13.2 to 15.1 seconds, against 1.0 second with \sys.
The GPU finishes the prefill in 0.11 seconds.
As a result, the TTFT with \sys is 1.1 seconds.

\subsection{Deep Dive \label{sec:eval-deep}}

\noindent\textbf{Sources of the gain.}
Table~\ref{tab:chain} separates the gain over \reins, the fastest existing scheme on 1\,GB and 8\,GB (Figure~\ref{fig:compare}), into the batching design of \sys and the matrix accelerator.
Without the accelerator, \vnni is already better than \reins by 4.8$\times$ and 8.1$\times$ on 1\,GB and 8\,GB, respectively.
This is because it batches both the scan and packing.
AMX then raises the throughput by another 5.0$\times$ and 5.2$\times$, respectively, because the vector units exhaust their compute capacity at small batches.
\sys needs both its design and the accelerator.
A CPU without AMX still gains from the design through \vnni.

\begin{table}[t]
\centering
\small
\caption{Where the gain over \reins comes from.\label{tab:chain}}
\begin{tabular*}{\linewidth}{@{\extracolsep{\fill}} l rr rr}
\toprule
 & \multicolumn{2}{c}{1\,GB} & \multicolumn{2}{c}{8\,GB} \\
\cmidrule(lr){2-3}\cmidrule(lr){4-5}
System & QPS & Ratio & QPS & Ratio \\
\midrule
\reins & 83.2 & 1.0$\times$ & 11.5 & 1.0$\times$ \\
\vnni & 403 & 4.8$\times$ & 93.8 & 8.1$\times$ \\
\sys & 2,013 & 24.2$\times$ & 492 & 42.6$\times$ \\
\bottomrule
\end{tabular*}
\end{table}

\begin{figure}[t]
\centering
\includegraphics[width=\linewidth]{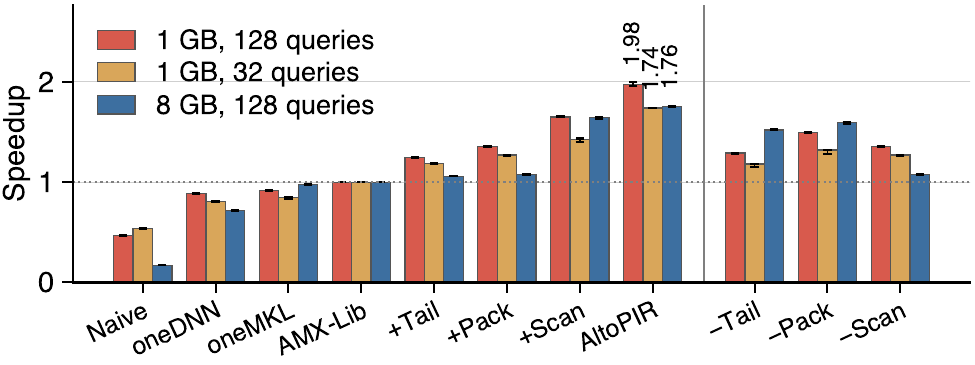}
\caption{Every stage adds to the speedup over \port. Bars right of the line remove one stage from +Scan.}
\label{fig:ablation}
\end{figure}

\noindent\textbf{Ablation study.}
To examine how much each mechanism contributes, we replace one stage of \port at a time with the corresponding stage of \sys, as Figure~\ref{fig:ablation} shows.
Every stage adds to the speedup.
On 1\,GB, the vectorized tail gives the largest step, 1.25$\times$, because it shortens the tail from 22.4\,ms to 5.5\,ms per batch.
On 8\,GB, the query-blocked scan gives the largest step, 1.52$\times$, because the scan takes 70\% of a batch of \port there.
Together, they make \sys faster than \port by 1.98$\times$ and 1.76$\times$ on 1\,GB and 8\,GB, respectively.
\sys raises the accelerator utilization from 22\% to 43\% on 1\,GB.

\noindent\textbf{Scan against library kernels.}
Table~\ref{tab:scans} compares the scan of \sys with four scans built on oneDNN and oneMKL, each run alone.
The scan of \sys is faster than the fastest library scan by 1.15$\times$ and 1.6$\times$ on 1\,GB and 8\,GB, respectively.
The table reports the fastest run of each library scan, which favors the libraries.
All scans keep the AMX unit busy for the same cycles.
The difference is that the library scans read the database from memory 1.7 to 4.7 times, while \sys reads it about once.
Therefore, the libraries are slower because of memory traffic rather than their use of the accelerator.

\begin{table}[t]
\centering
\small
\caption{Isolated scans of 128 queries. \sys is the fastest and reads the database about once. DRAM reads are relative to the database size and L2 misses to \sys.\label{tab:scans}}
\setlength{\tabcolsep}{3pt}
\begin{tabular}{l rr rr r}
\toprule
 & \multicolumn{2}{c}{Time (ms)} & \multicolumn{2}{c}{DRAM reads} & L2 \\
\cmidrule(lr){2-3}\cmidrule(lr){4-5}
Scan & 1\,GB & 8\,GB & 1\,GB & 8\,GB & misses \\
\midrule
oneDNN & 39.9 & 377 & 2.16$\times$ & 2.65$\times$ & 4.18$\times$ \\
oneMKL & 33.5 & 363 & 2.56$\times$ & 3.02$\times$ & 2.89$\times$ \\
brgemm & 47.6 & 357 & 3.39$\times$ & 4.68$\times$ & 6.38$\times$ \\
brgemm, \sys order & 37.3 & 359 & 1.74$\times$ & 2.49$\times$ & 4.91$\times$ \\
\sys, spilled sums & 31.0 & 236 & 1.97$\times$ & 1.88$\times$ & 1.53$\times$ \\
\sys, no prefetch & 31.2 & 242 & -- & -- & 1.01$\times$ \\
\sys & 29.1 & 224 & 1.19$\times$ & 1.17$\times$ & 1.00$\times$ \\
\bottomrule
\end{tabular}
\end{table}

\begin{figure}[t]
\centering
\subfloat[Operand tiles held fixed]{\includegraphics[width=0.48\linewidth]{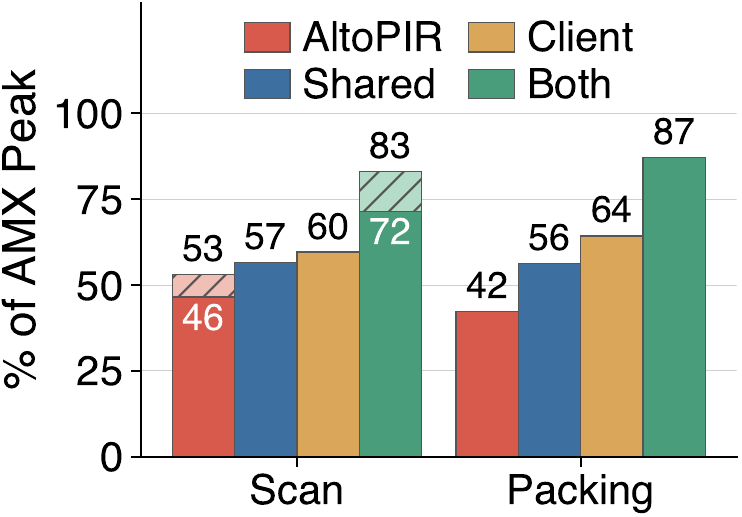}}\hfill
\subfloat[Block length of the scan]{\includegraphics[width=0.48\linewidth]{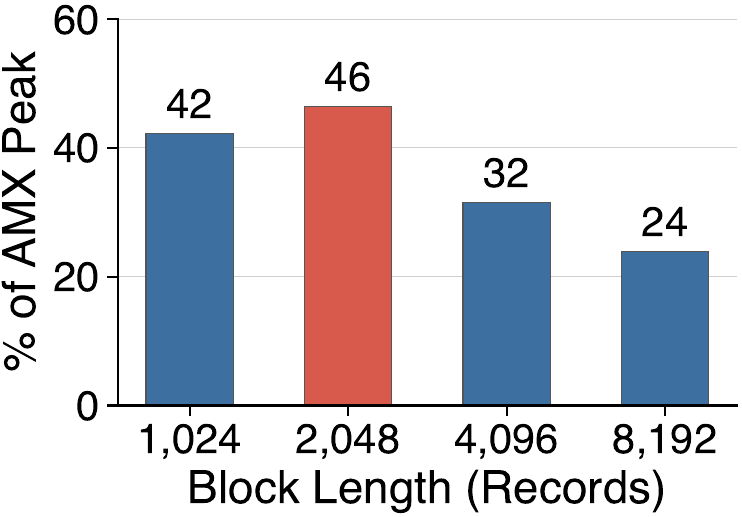}}
\caption{Loading operand tiles from the L2 cache limits the scan and packing of \sys (1\,GB, 128 queries). (a) Shared, Client, and Both hold those tiles at one address, and hatches skip the recombination. (b) The block length that \sys chooses is the fastest.}
\label{fig:feed-probe}
\end{figure}

\begin{figure}[t]
\centering
\includegraphics[width=\linewidth]{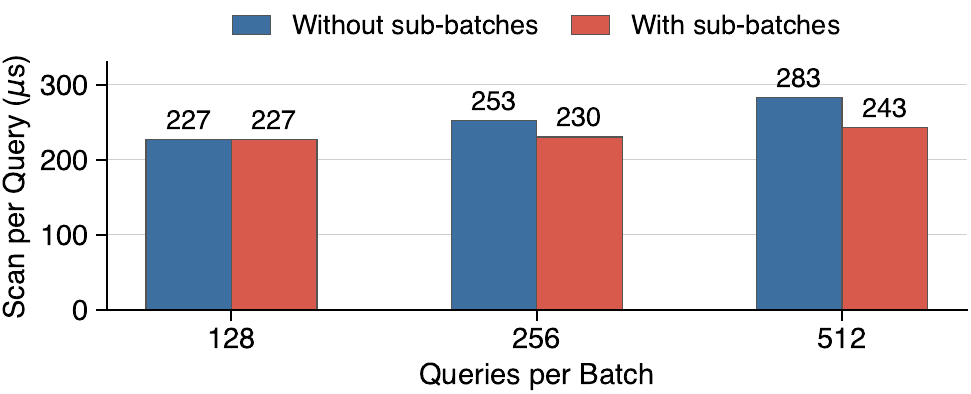}
\caption{Impact of sub-batches. Sub-batches keep the scan time per query nearly flat as the batch grows.}
\label{fig:place}
\end{figure}

\noindent\textbf{Feeding the accelerator.}
We rerun the kernels of \sys with operand tiles held at fixed addresses, which removes the cost of loading them but gives wrong results.
Figure~\ref{fig:feed-probe}~(a) shows that the scan of \sys reaches 46\% of the peak AMX rate and packing 42\%.
Fixed operands raise them to 72\% and 87\%.
This means that loading tiles from the L2 cache explains most of the packing gap and about half of the scan gap.
Figure~\ref{fig:feed-probe}~(b) shows that the block length of \sys gives the highest scan rate, because shorter blocks recombine more often and longer ones overflow the L2 cache.
Sub-batches keep this block length for batches of more than 128 queries.
With them, the scan time per query grows by only 7\% from 128 to 512 queries per batch, as Figure~\ref{fig:place} shows.
Without sub-batches, it grows by 25\%, because the block length shrinks to 512 records.

\begin{figure}[t]
\centering
\includegraphics[width=\linewidth]{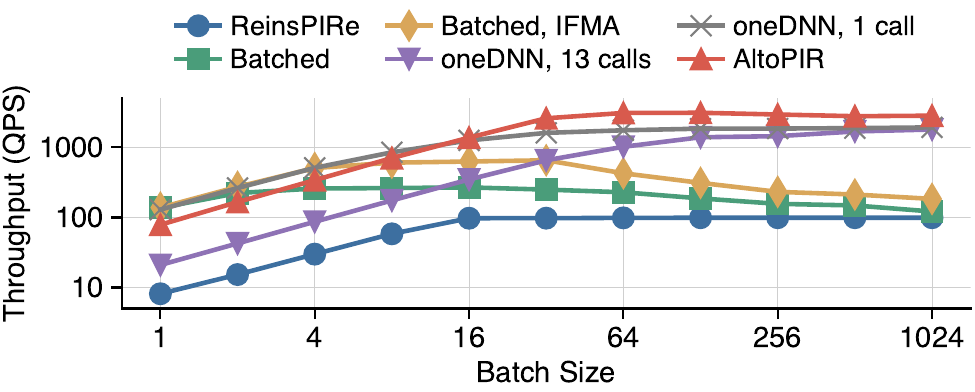}
\caption{The query-blocked scan also speeds up the online step of \reins. \reins is its own data flow, one query per thread, and Batched uses AVX-512.}
\label{fig:reins}
\end{figure}

\noindent\textbf{Generality to other schemes.}
To check that the techniques of \sys are not specific to \sandwich, we apply the query-blocked scan to the dominant online step of \reins.
In its code, this step multiplies 16-bit entries by 52-bit query values modulo $2^{52}$.
Figure~\ref{fig:reins} compares each data flow at its best batch size.
This scan achieves higher throughput than the data flow of \reins by 31.5$\times$ and than the best data flow with oneDNN products by 1.6$\times$.
With more byte limbs per query value, it is slower than oneDNN below 16 queries per batch.
A complete \reins would still be limited by its per-client steps to 542\,QPS.
These results show that the scan of \sys also applies to other hintless schemes.
Their per-client steps decide how much of this gain they keep.

\begin{figure}[t]
\centering
\includegraphics[width=\linewidth]{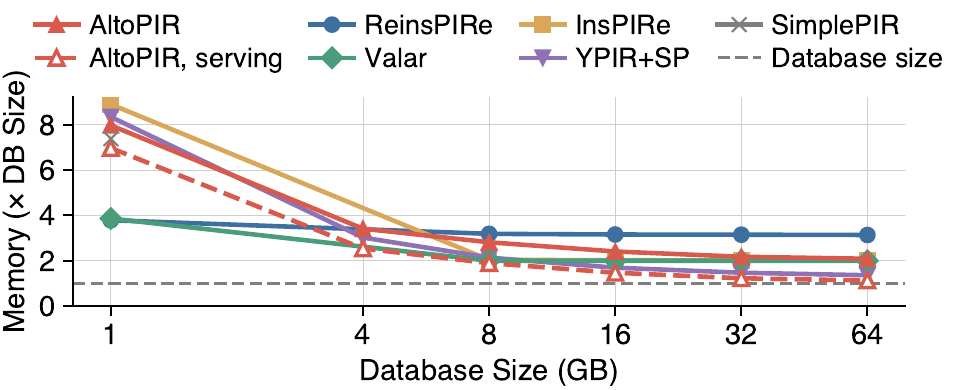}
\caption{Memory divided by the database size. From 4\,GB on, two copies of the database make up most of the peak of \sys, and it frees one of them after setup (dashed).}
\label{fig:memory}
\end{figure}

\noindent\textbf{Memory usage.}
Figure~\ref{fig:memory} shows that the ratio of the peak memory of \sys to the database size falls from 8.0 at 1\,GB to 2.1 at 64\,GB.
This is because part of its state has a fixed size, such as the 1\,GB packing matrix for 32\,KB records.
From 4\,GB on, two copies of the database during setup make up most of the peak, and \sys frees one of them once setup ends.
At 64\,GB, this ratio is close to those of the existing schemes.

\noindent\textbf{Database updates.}
Table~\ref{tab:updates} shows the time to apply record updates with the rank-one update of the hint.
Changing one record takes 2.6\,s on both 1\,GB and 8\,GB databases, far less than a full setup.
This is because each changed record adds only a hint update of about 18\,ms and a rewrite of the two database copies of 1 to 3\,ms.
The rest of the state takes a fixed 2.4\,s to recompute.
Afterward, the state and the responses equal those of a fresh instance of \sys set up on the updated database.
In practice, a service can apply small updates within seconds.

\begin{table}[t]
\centering
\small
\caption{Time to apply record updates. Updating a few records costs much less than building the offline state again.\label{tab:updates}}
\begin{tabular*}{\linewidth}{@{\extracolsep{\fill}} l r r}
\toprule
 & 1\,GB & 8\,GB \\
\midrule
1 changed record & 2.6\,s & 2.6\,s \\
64 changed records & 3.4\,s & 3.6\,s \\
Full setup & 13.1\,s & 89\,s \\
\bottomrule
\end{tabular*}
\end{table}

\begin{figure}[t]
\centering
\includegraphics[width=\linewidth]{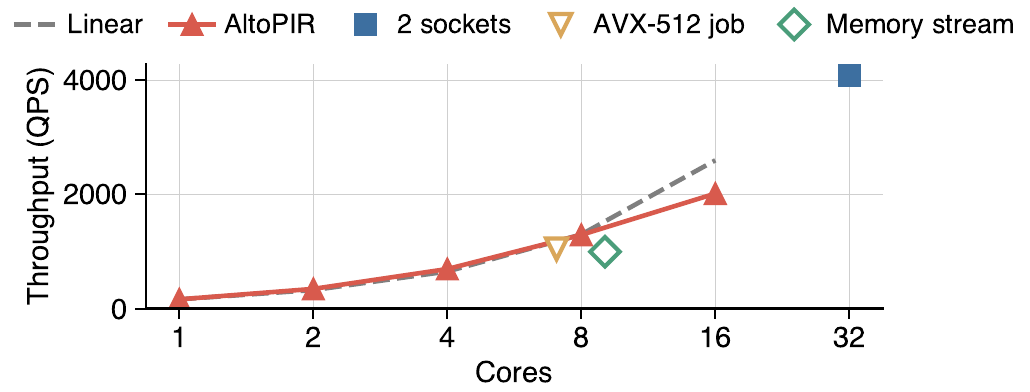}
\caption{Throughput grows nearly in proportion to the number of cores up to eight. Separate points show two sockets and eight cores beside another job.}
\label{fig:cores}
\end{figure}

\noindent\textbf{Core scaling.}
Figure~\ref{fig:cores} shows that \sys scales almost linearly up to eight cores and reaches 0.78 of linear scaling at 16 cores.
This is because the AMX clock drops from 3.40\,GHz with one busy core to 2.49\,GHz with 16.
A second socket with its own copy of the database raises the throughput by 2.03$\times$.
When the other eight cores run an AVX-512 job or use 135\,GB/s of memory bandwidth, \sys on eight cores keeps 81\% or 78\% of its throughput.
Therefore, \sys scales with the number of sockets and can run beside other jobs.

\begin{figure}[t]
\centering
\subfloat[Block length of the scan]{\includegraphics[width=0.48\linewidth]{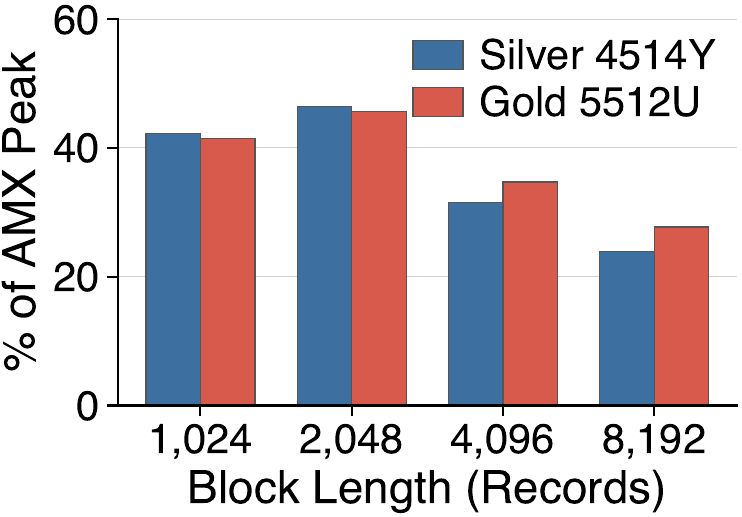}}\hfill
\subfloat[Throughput on 1\,GB]{\includegraphics[width=0.48\linewidth]{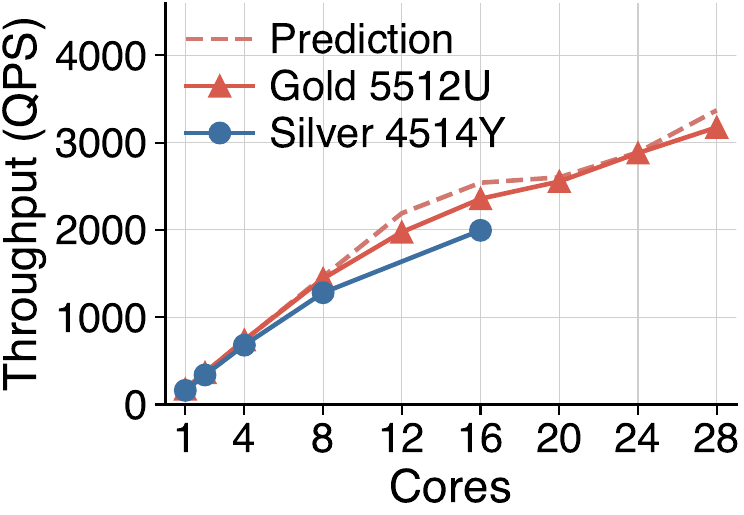}}
\caption{\sys is robust to another CPU.}
\label{fig:sku}
\end{figure}

\noindent\textbf{Robustness on another CPU.}
We also run \sys on a CloudLab server~\cite{cloudlab} with an Intel Xeon Gold 5512U, a CPU of the same generation as the main one but with 28 cores.
With the same 2\,MB L2 cache per core, the rules of \sys give the same block length.
Figure~\ref{fig:sku}~(a) shows that this block length again gives the highest scan rate.
Figure~\ref{fig:sku}~(b) shows that, at every core count, the throughput stays within 10\% of the throughput that we predicted from the main CPU before the runs.
From 16 to 28 cores, the throughput grows by only 1.35$\times$, because the AMX clock falls from 3.18\,GHz to 2.39\,GHz when all 28 cores are busy.
In summary, \sys also performs as predicted on another CPU, and its throughput scales with the cores and the clock of the accelerator.

\section{Related Work \label{relatedwork}}

\noindent\textbf{PIR schemes and batching.}
Lattice-based schemes made PIR practical~\cite{xpir,sealpir,onionpir,spiral,respire}, and hint-based schemes cut the online computation of the server with a digest that each client stores~\cite{simplepir,frodopir,piano}.
Hintless schemes instead move this digest to the server~\cite{hintlesspir,ypir,inspire,reinspire} or use no hint~\cite{kspir,whispir,via}.
Unlike batch PIR for one client~\cite{batchcodes,sealpir,vbpir,pirana}, cross-client batching answers independent clients with one product over the database~\cite{luekspir,ypir}, like the shared scans of database systems~\cite{cooperativescans,qpipe,shareddb}.

\noindent\textbf{Accelerators and matrix arithmetic.}
Other PIR systems, including multi-server ones~\cite{guentherpir,lamgpupir}, add GPUs~\cite{gpir,vipir,dpir,shiftpir}, computational storage~\cite{inspireisca,smartpir}, or a dedicated accelerator~\cite{ive} to every server.
\sandwich~\cite{sandwichpir}, whose scheme \sys follows, runs the scan and packing on GPU tensor cores, while \sys runs on CPU matrix accelerators that hold few operands and share their cores with request handling.
GPIR~\cite{gpir} also finds that batching overflows the L2 cache of a GPU and fuses the operations of each overflowing step into one kernel.
\sys has a smaller budget, because AMX needs its operands in the L1 cache to run at its full rate.
IVE~\cite{ive} finds that batching leaves the steps without a shared operand bound by memory, whereas packing in \sys shares the packing matrix across clients.
Matrix units also run mixture-of-experts inference~\cite{ktransformers}, and wide products built from 8-bit products combine their partial products in a separate step~\cite{sandwichpir,ozakiint8,hoevenamx,kouyaamx} or inside one GPU kernel~\cite{emugemm,vipir}, whereas \sys combines them after every block in the L2 cache.
GEMM libraries block one product for the caches~\cite{gotoblas,blis,brgemm,onednn}, while \sys sets the blocks of the scan and packing within one cache budget.

\section{Conclusion \label{conclusion}}

We presented \sys, the first PIR system built on the matrix accelerators inside server CPUs.
\sys splits encrypted values into bytes for the accelerator.
It reads the database once for many queries, keeps each core fed from its caches, and handles each request in place with vector instructions.
On a commodity server without added devices, it outperforms the fastest existing hintless scheme by 24.2$\times$ to 50.1$\times$ on databases of 1\,GB to 64\,GB.
We believe that the accelerators built into server CPUs can make PIR, and other privacy systems whose computation reduces to integer matrix products, practical on commodity servers.

\label{lastpage1}
\section*{Acknowledgement}
This research was sponsored by the National Research Foundation of Korea (NRF) grants funded by the Ministry of Science and ICT (No. RS-2025-00522990). Gyuyeong Kim is the corresponding author.

\bibliographystyle{plain}
\begingroup
\sloppy
\setlength{\emergencystretch}{3em}
\bibliography{./gykim,./altopir}
\endgroup

\label{lastpage2}
\end{document}